\documentclass[12pt]{article}
\usepackage[utf8]{inputenc}

\usepackage{amsmath,amsfonts,graphicx}
\usepackage[square,numbers,sort&compress]{natbib}
\usepackage{fullpage}
\usepackage{lineno}
\usepackage{xcolor}
\usepackage{pdflscape}
\usepackage{hyperref}
\usepackage{multicol,multirow}
\usepackage{authblk}

\newcommand{\hl}{}                    

\title{Joint economic and epidemiological modelling of alternative pandemic response strategies }

\author[1,2]{Michael J. Plank}
\author[1]{Megan Sushames}
\author[1]{Timothy Fisher-Taylor}
\author[3]{Amin Afshari}
\author[4]{Robin N. Thompson}
\author[5]{Amy Hurford}
\author[2,6,7]{Shaun C. Hendy}

\affil[1]{School of Mathematics and Statistics, University of Canterbury, Christchurch, New Zealand}
\affil[2]{Te P\=unaha Matatini, Auckland, New Zealand}
\affil[3]{Department of Biological Sciences, Simon Fraser University, British Colombia, Canada}
\affil[4]{Mathematical Institute, University of Oxford, Oxford, United Kingdom}
\affil[5]{Biology Department and Department of Mathematics and Statistics, Memorial University of Newfoundland, St John's, Newfoundland and Labrador, Canada}
\affil[6]{School of Science in Society, Te Herenga Waka - Victoria University of Wellington, New Zealand}
\affil[7]{Toha Network Limited, Gisborne, New Zealand}

\date{}

\begin{document}

\maketitle

\newpage


\begin{abstract}
In an emerging pandemic, policymakers need to make decisions with limited information and requiring trade-offs between the health impact of the pandemic and the economic costs of the response. Most mathematical models have focused on direct health impacts, neglecting the economic costs of control measures. Here, we introduce a framework that captures both health and economic costs and compare the expected aggregate costs of alternative strategies across a range of epidemiological and economic parameters. We find that for diseases with low severity, mitigation tends to be the most cost-effective option. For more severe diseases, suppression tends to be most cost-effective if the basic reproduction number $R_0$ is relatively low, while elimination tends to be more cost-effective if $R_0$ is high. We use the example of New Zealand's response to the Covid-19 pandemic in 2020 to anchor our framework to a real-world case study. We find that parameter estimates for Covid-19 in New Zealand put it close to or above the threshold at which elimination becomes more cost-effective than mitigation. We conclude that our proposed framework holds promise as a decision-support tool for pandemic threats, with further work needed to account for population heterogeneity and other factors relevant to decision-making.

\end{abstract}

\newpage

\section{Introduction}

Emerging pandemics require timely selection of public health strategies under conditions of high uncertainty and limited data. This was exemplified during the Covid-19 pandemic, when different countries took very different approaches to responding to the pandemic. When faced with an emerging pandemic or infectious disease threat, mathematical models are a useful tool to support impact assessment \cite{reed2013novel,shearer2021development} and to inform evaluation of controllability  \cite{fraser2004factors,mccaw2014pandemic} and strategy selection \cite{french2025preparing}.

Pandemic responses can be broadly categorised into three types of strategy: mitigation, suppression and elimination \cite{baker2023covid}. 
A mitigation strategy aims to slow transmission and reduce the total number of people infected and the peak demand on healthcare systems. This is sometimes referred to as a ``flatten the curve'' strategy. A suppression strategy aims to keep prevalence low by maintaining the effective reproduction number at or below $1$ on average. An elimination strategy aims to reduce the number of incident infections to zero \cite{wu2021aggressive}.

Suppression and elimination strategies are typically temporary strategies as they do not remove the threat of an epidemic given a susceptible population, while mitigation might be sustained throughout an epidemic. During the Covid-19 pandemic, suppression and elimination strategies were employed in some countries to maintain infections at a low level until vaccines became available and public health and social measures could be relaxed. Local elimination does not usually mean that a pathogen is eradicated; the pathogen may be reintroduced from a range of sources \cite{metcalf2021challenges}, including importation from another location \cite{douglas2021real}, relapse of a previously identified case \cite{lee2017recrudescence}, or zoonotic spillover \cite{lai2016global,dudas2018mers}.
Furthermore, the feasibility of alternative strategies will be situation-dependent: elimination may not be feasible in jurisdictions that have large volumes of uncontrolled border traffic, extensive land borders, or under-resourced public health systems \cite{martignoni2024sars}.

Previous studies have attempted to calculate the optimal way to control an epidemic, typically by optimising a given objective function subject to certain constraints. Examples include minimising the cumulative number of infections subject to constraints on the intensity, duration or cost of control \cite{hansen2011optimal,bliman2021best,britton2023optimal}; and minimising the cumulative cost of controls while preventing the number of severe cases from exceeding healthcare capacity \cite{godara2021control}. There are also studies that model the trade-off between the costs of control measures and the costs of infections, either from the point of view of a central planner or self-interested individuals \cite{reluga2010game,fenichel2011adaptive,fenichel2013economic,toxvaerd2019rational}. These reveal that, in both cases, the optimal response involves a reduction in contact rates when disease prevalence is high. However, the endogenous behavioural response of self-interested individuals is typically suboptimal from a central planner's perspective, because individuals do not account for the effect of their own behaviour on the risk to others  \cite{reluga2010game,gersovitz2004economical,eichenbaum2021macroeconomics,auld2025economics}.  Models for the dynamics of the behavioural response to an epidemic have also been considered. For example, ``information dynamics'' or ``opinion dynamics'' models treat the strength of the behavioural response as a dynamic variable that adjusts according to current infection prevalence either at the population-level \cite{ryan2024behaviour,plank2026complex} or in a local area of the social contact network \cite{funk2009spread,chang2025impact}. 

Most studies concerned with optimal epidemic control consider mitigation responses, in the sense that they require that the herd immunity threshold is eventually reached so that there is no further risk of resurgent epidemic activity and therefore no further need for control measures. Comparatively little attention has been paid to comparing the aggregate cost of a mitigation response with that of an elimination of suppression response in different epidemiological contexts. 
Martignoni et al. \cite{martignoni2024sars} argued that there are three essential considerations in determining whether elimination is the preferred strategy: healthcare capacity, epidemiological feasibility and cost-effectiveness.

Here, we propose a novel framework that allows the expected costs of mitigation, suppression and elimination strategies to be compared across a range of scenarios. Our model for mitigation shares similarities with those described in \cite{toxvaerd2019rational,eichenbaum2021macroeconomics,auld2025economics}. We formulate a cost function that comprises the direct costs of the epidemic (due to mortality and morbidity) and the costs of public health and social measures (PHSMs) or behavioural changes that occur in response to the epidemic. In keeping with the economic literature, we model interventions (whether mandated centrally or taken voluntarily) as a costly reduction in potentially transmissive contacts \cite{auld2025economics}.
We calculate the control policy that minimises the aggregate cost, as well as the endogenous response that arises as a result of optimal individual behaviour in the absence of centrally imposed controls. This allows us to estimate the net benefit of intervention against a no-intervention benchmark that takes account of the expected behavioural response.   

We also develop models for the cost of a suppression and an elimination strategy. Mathematically, we distinguish between these by defining suppression to be a reduction in contact rates sufficient to reduce prevalence to a low level and then maintain the effective reproduction number close to $1$ in a fully susceptible population. On the other hand, we interpret elimination as using strict border controls to minimise the number of imported infections, and only using community control measures when required to eliminate border-related outbreaks. The distinction between elimination and suppression is not always clear-cut. For example, a suppression strategy may inadvertently result in elimination of the pathogen; whilst an elimination strategy may become indistinguishable from suppression if the number of imported cases is high. Alternative definitions are possible \cite{wu2021aggressive} but the distinction between primarily community control versus primarily border control is useful for the present work. 

We use the example of New Zealand's response to Covid-19 in 2020, which used both border restrictions and community controls of varying degrees of stringency \cite{baker2020successful,steyn2021managing,hendy2021mathematical}, to inform parameter estimates relating to the costs of control measures. However, our {\hl model is more generally applicable to pandemics of respiratory pathogens}: we compare the costs of alternative strategies across a range of scenarios representing combinations of disease transmissibility and severity. We also consider how the frequency of border-related outbreaks (i.e. outbreaks initiated by imported infections), the ability of surveillance to detect outbreaks in a timely way, and the effectiveness of test-trace-isolate (TTI) measures affect cost-effectiveness. 

{\hl Our aim is to provide a framework that can support policy-makers in planning the response to a future pandemic in a range of epidemiological and economic scenarios. We use a relatively simple model that captures broad-scale effects, rather than granular details, so that our framework is adaptable to different contexts. The framework could form the starting point for a more detailed pathogen- and country-specific analysis in the event of a future emerging infectious disease threat.}

\section{Methods}

\subsection{Transmission model}

Consider a simple compartment-based model for the number of susceptible individuals $S(t)$ and infectious individuals $I(t)$ in a closed, homogeneous population of size $N$:
\begin{align}
    \frac{dS}{dt} &= -\beta a(t)^2 S\frac{I}{N} \label{eq:dSdt} \\
    \frac{dI}{dt} &= \beta a(t)^2 S\frac{I}{N} - \gamma I \label{eq:dIdt} 
\end{align}
where $a(t)\in[0,1]$ is the ``activity level'' at time $t$ relative to baseline. When there is no policy intervention or behavioural change, we have $a(t)=1$; smaller values of $a(t)$ represent larger reductions in activity. The force of infection in Eqs. \eqref{eq:dSdt}--\eqref{eq:dIdt} is proportional to $a(t)^2$ because this represents the product of the activity levels of individuals on both sides of the susceptible-infectious pair (termed `quadratic matching' in \cite{diamond1979equilibrium}). When there is no policy intervention, the basic reproduction number $R_0$ for this model is $\beta/\gamma$.

Suppose there is an average cost per person $k>0$ of being infected. {\hl This may be interpreted as the cost that an individual is willing to pay to avoid infection, or that a central planner is willing to pay to avert one infection. In general, there may be different factors contributing to $k$, for example risk of death or reduction in quality in life, financial costs of healthcare, and lost earnings while sick. For modelling purposes, we abstract this complexity into a single parameter.  Nevertheless, in Supplementary Material sec. S3, we describe two approaches for estimating $k$. The first approach estimates $k$ as a revealed cost based on the observed magnitude of endogenous behavioural change. The second approach combines estimates of various costs including: (i) direct financial costs; and (ii) mortality and morbidity costs, which we convert to a monetary value using an estimate of the value of a disability-adjusted life year (DALY) \cite{daroudi2021cost}. }

Suppose there is also a cost to reducing activity (whether voluntarily or due to government restrictions), so that the cost per person per unit time is $c_1\left( 1-a(t)\right)+ c_2\left(1-a(t)\right)^2$ for some parameters $c_1,c_2\ge 0$. This means that the cost per unit time of reducing activity is increasing and convex with respect to the magnitude of the reduction, $1-a(t)$. Hence, whenever $c_2>0$, the marginal cost of further reductions in activity increases as activity is reduced. Methods for estimating $c_1$ and $c_2$ empirically are described in Supplementary Material sec. S3.

\subsection{Centralised mitigation response} \label{sec:cent}

In this section, we formulate the optimal control problem for a central planner who seeks to determine the time-dependent activity level $a(t)$ that minimises the aggregate cost. We consider the simplest version of the problem in which the central planner has no way to distinguish which disease compartment an individual is in, so the activity level $a(t)$ applies to the whole population. Ignoring discounting and expressing the objective function as the per capita cost, this leads to the following optimisation problem:
\begin{equation} \label{eq:central_planner}
    \min_{a(t)} \left[ k\left(1-\frac{S_\infty}{N}\right) + c_1\int_0^\infty \left(1-a(t)\right) dt + c_2\int_0^\infty \left(1-a(t)\right)^2 dt \right]
\end{equation}
where $S_\infty=\lim_{t\to \infty}S(t)$ is the number of people who remain uninfected at the end of the epidemic. Because of the infinite time horizon in Eq. \eqref{eq:central_planner}, activity levels must eventually return to normal (i.e. $a(t)=1$ for sufficiently large $t$). This implies that the cumulative number of infections must eventually exceed the herd immunity threshold, $1-1/R_0$. Therefore, the solution to Eq. \ref{eq:central_planner} represents the optimal centralised mitigation response and is referred to as `socially optimal' because it results in the lowest net cost for all individuals in the population (under the assumption of a homogeneous population). In practice, we solve Eq. \eqref{eq:central_planner} with a fixed time horizon $T$ that is sufficiently large that the optimally controlled epidemic has ended before time $T$. We use a direct numerical optimisation method \cite{montes2025exploring} (for details see Supplementary Material sec. 1.1). 

It should be noted that Eq. \eqref{eq:central_planner} does not account for additional health and economic costs that may arise if healthcare systems are overwhelmed. This could be modelled by imposing a constraint on the maximum infection prevalence $I(t)$, or by adding escalating costs into Eq. \eqref{eq:central_planner} when prevalence exceeds the level corresponding to healthcare capacity. We leave this extension for future work, noting that the existence of a single threshold at which the healthcare system becomes overwhelmed is itself a major simplification \cite{fong2024nhs}.

\subsection{Decentralised mitigation response} \label{sec:decent}

In the absence of centrally imposed controls, rational self-interested individuals will seek to choose their own time-dependent activity level $a(t)$ in such a way as to minimise their own cost. In general, the optimal activity level for a given individual will depend on the activity levels of others. A Nash equilibrium for this system is a situation where: (i) all individuals have the same time-dependent activity level $a(t)$; and (ii) any individual choosing a different activity level pays a higher cost \cite{reluga2010game}. In Supplementary Material sec 1.2, we derive an equation for a Nash equilibrium and provide the numerical method used to calculate it. Again, we consider the simplest version of the problem in which individuals do not modify their behaviour according to which disease compartment they are in (see Supplementary Material sec. 1.4 for a more general, state-dependent version of the model).

In addition to the simplifying assumptions of a homogeneous compartment-based model, the Nash equilibrium assumes that individuals have perfect knowledge of the current state and future dynamics of the epidemic and the cost function, and are able to instantaneously adjust their activity level to the rational solution. This is clearly a major abstraction and we would not expect this to be perfectly realised in a real epidemic (in the same way that we would not expect the centralised solution to be perfectly attainable). Nevertheless, it serves as a guide to the expected magnitude of the endogenous behavioural response to an epidemic in the absence of centrally imposed interventions. It captures the fact that if the bulk of the population's activity level is above the Nash equilibrium at a given point in time, the rational response for a representative individual is to decrease activity, and vice versa.

\subsection{Suppression and elimination}

In an ordinary differential equation model, the infection prevalence $I(t)$ never reaches zero. This means that, if activity levels return to normal ($a(t)=1$) and the susceptible population is sufficiently large ($S(t)/N>1/R_0$), an epidemic will always occur. Hence, the optimisation problems defined in Sections \ref{sec:cent}--\ref{sec:decent} will never produce solutions resulting in elimination.

Here, we develop a simplified model for the costs of elimination and suppression strategies. In both cases, we assume there is a fixed, known time horizon $T$ at which a completely effective vaccine or therapy becomes widely available. We assume that this reduces all infection costs after time $T$ to zero (either by preventing transmission entirely or by reducing the severity of disease to a negligible level), effectively ending the epidemic and the need for any further controls. This is a major simplification: in reality, a vaccine or therapy will not be 100\% effective nor achieve 100\% coverage and will take time to roll out the population. However, it is a useful approximation provided the vaccine or therapy can reasonably be expected to reduce the clinical severity of infection in vulnerable groups to manageable levels.

{\bf Suppression.} To model suppression, we assume that infection prevalence starts at a low level, and community controls are imposed that maintain the effective reproduction number $R_\mathrm{eff}$ equal to $1$ in a fully susceptible population. We assume that, because infection prevalence remains low, TTI measures can be used and the effect of these is to multiply the transmission rate by a constant factor $\alpha\in[0,1]$ (in addition to and independent of the population-wide activity level reduction). Then, the value of $a(t)$ required to reduce $R_\mathrm{eff}$ to 1 is $a_\mathrm{sup} = 1/\sqrt{\alpha R_0}$. We assume that $a(t)=a_\mathrm{sup}$ for all $t\in[0,T]$ and that prevalence is kept sufficiently low that the cost of infections is negligible. We also assume that the cost of TTI measures is negligible in comparison to blanket population-level community controls. The expected per capita cost of a suppression response is therefore modelled as
\begin{equation}  \label{eq:sup_cost}
C_\mathrm{sup} = \left[ c_1\left(1-a_\mathrm{sup}\right) + c_2\left(1-a_\mathrm{sup} \right)^2 \right] T
\end{equation}
Note that this corresponds to what is sometimes called `strict suppression' or `tight suppression' \cite{blakely2021association} in that it requires prevalence to be kept to a very low level, in contrast to `loose suppression' which tolerates a higher prevalence and hence allows some immunity to build up in the population. 

{\bf Elimination.} To model elimination, we assume that strict border controls are imposed continuously, at a constant cost $b$ per unit time. This reduces the number of imported infections, but due to imperfect control \cite{grout2021failures} some transmission from imported infections into the community is still expected to occur. {\hl In reality,} some border-related outbreaks may self-extinguish without controls due to stochastic effects, {\hl which are not included in the model}; others that are detected sufficiently early may be eliminated by TTI measures \cite{douglas2021real}. {\hl We do not explicitly model these because they do not require introduction of PHSMs and, therefore, have negligible cost. Instead, we only consider those outbreaks that grow undetected to a size where they require community controls.} We assume that such outbreaks occur at a frequency $r$ per unit time and that each such outbreak grows exponentially at rate $\beta-\gamma$ for some time period $t_\mathrm{det}$ until being detected, representing an uncontrolled outbreak in a fully susceptible population. We assume that, once the outbreak is detected, TTI measures multiply the transmission rate by a constant factor of $\alpha$ and community controls are introduced that reduce activity levels to $a_\mathrm{elim}$ (further reducing transmission by a factor of $a_\mathrm{elim}^2$) until the outbreak is eliminated. Under these assumptions, if the initial prevalence is $I_0=1$, then the prevalence when the outbreak is first detected is $I_\mathrm{det}=e^{(\beta-\gamma)t_\mathrm{det}}$. The duration of controls needed to reduce prevalence below $1$ (at which time we assume the outbreak is eliminated) is 
    \begin{equation}
        t_c = \frac{\beta-\gamma}{\gamma-\beta \alpha a_\mathrm{elim}^2}t_\mathrm{det}
    \end{equation}
    The cost of doing this is $t_c( c_1(1-a_\mathrm{elim})+c_2(1-a_\mathrm{elim})^2  )$ and it is straightforward to find the optimal value of the constant $a_\mathrm{elim}\in[0, a_\mathrm{sup})$ to minimise this. Finally, we assume that border-related outbreaks occur non-concurrently, and are controlled by applying control measures to some fraction $p_\mathrm{cont}$ of the economy, representing the city or region in which the outbreak occurs. 
    
    This is a highly simplified deterministic model for sporadic outbreaks that {\hl ignores all stochastic variability in the transmission, detection and control processes}. We would not expect this model to accurately describe any given outbreak, and ignoring stochastic effects in the epidemic decay phase means it may overestimate outbreak duration. We therefore interpret it as an approximate model for the expected cost averaged over a large number of outbreaks occurring under comparable conditions. 
    
    As above, we assume that the cumulative number of infections from sporadic outbreaks is small enough that their aggregate cost is negligible (see Supplementary Material sec. 2 for details). Under these assumptions, the expected per capita cost of an elimination strategy is
    \begin{equation} \label{eq:elim_cost}
      C_\mathrm{elim} = \left[ \underbrace{b\vphantom{\frac{a_a}{b_b}}}_\textrm{border control cost} +  \underbrace{r p_\mathrm{cont}  t_\mathrm{det}   (\beta-\gamma) \frac{ c_1(1-a_\mathrm{elim})+c_2(1-a_\mathrm{elim})^2  }{\gamma-\beta \alpha a_\mathrm{elim}^2}}_\textrm{cost of eliminating small outbreaks} \right] T
    \end{equation}
    For the purposes of visualisation, we also calculate the time-dependent activity level $a(t)$ and the cumulative cost of elimination assuming that border-related outbreaks occur deterministically every $1/r$ days. In this situation, the activity level alternates between non-outbreak periods of duration $1/r-t_c$ when $a(t)=1$, and outbreak periods of duration $t_c$ when $a(t)=a_\mathrm{elim}$. During non-outbreak periods, the cost per unit time is $b$, while during outbreak periods, it is $b+c_1(1-a_\mathrm{elim})+c_2(1-a_\mathrm{elim})^2$.

    Note that neither the suppression nor the elimination model accounts for the cost of controlling a larger initial outbreak, such as the Covid-19 outbreaks that occurred in many countries in February-March 2020. If the decision between elimination/suppression and mitigation is being considered when an outbreak is already underway, this should be factored in by adding the estimated cost of controlling the outbreak to the models for the cost of suppression or elimination.

\subsection{Parameter estimates for Covid-19 in New Zealand in 2020} \label{sec:params}

We present model results for a range of values of the basic reproduction number $R_0$ and the cost per infection $k$. We use fixed values for the control cost parameters $c_1$ and $c_2$ (see below). However, note that the optimisation problems representing the centralised and decentralised mitigation response depend only the ratios $c_1/k$ and $c_2/k$. Therefore, increasing $c_1$ and $c_2$ by a factor of $\phi$ is equivalent to reducing $k$ by a factor of $\phi$ and increasing the costs of all response types by a factor of $\phi$. 

Our results therefore encompass a wide variety of situations. However, to anchor the results in a real-world context, we use the Covid-19 pandemic in New Zealand as a case study for our model. New Zealand adopted an elimination strategy in response to Covid-19 in 2020-21, using strong border controls and relying on stringent PHSMs to eliminate the virus when border incursions occurred \cite{baker2023covid,mccaw2022role}. Using estimates of the costs and effectiveness of the different levels of PHSMs used in New Zealand \cite{TreasuryEconomicUpdateSept2020,hendy2021mathematical,CBEHendy2025}, we estimated the contact reduction cost coefficients $c_1$ and $c_2$ to be \$46 per person per day and \$36 per person per day respectively, and the border closure cost to be \$33.8 million per day (all costs are in New Zealand dollars). Using two different methods, we estimated the average cost $k$ per infection with {\hl severe acute respiratory syndrome coronavirus 2 (SARS-CoV-2)} to be between \$6500 and \$12000 (see Supplementary Material sec. S3 for details). 

These values are not intended to be definitive or universally applicable to other jurisdictions or other pandemics. Instead, they serve as a guide to the approximate region of parameter space that was relevant for the Covid-19 pandemic, and to inform expectations in future pandemic events. Precise parameter values for a specific emerging infectious disease threat will depend on characteristics of the pathogen, population, public health system, and local economy. {\hl In general, the parameter $k$ is likely to depend on both healthcare costs and morbidity and mortality costs, and is therefore not only a pathogen-specific parameter, but also captures a combination of pathogen and economic properties and will vary between countries.}

We investigate model sensitivity to values for the frequency and size of border-related outbreaks ($r$ and $t_\mathrm{det}$) and the TTI transmission multiplier $\alpha$. We run simulations for a population of size $N=5$ million, and a disease with an average infectious period of $1/\gamma=$ 5 days. All parameter values used in the simulations are shown in Table \ref{tab:params}. Code to reproduce the results in this article is available at \cite{github-repo}. All computations were run in {\em Matlab R2022b}.

 \begin{table}[]
     \centering
     \begin{tabular}{p{11cm}l}
     \hline
        Parameter  & Value  \\
        \hline
        Transmission rate & $\beta= 0.25$ -- $0.7$ day$^{-1}$ \\
        Recovery rate & $\gamma=1/5$ day$^{-1}$ \\
        Population size & $N=5\times 10^6$ \\
        Average cost per infection & $k =$ \$1000 -- \$15000\\
        Contact reduction cost per person per unit time (linear)  & $c_1=$ \$46 day$^{-1}$ \\
        Contact reduction cost per person per unit time (quadratic) & $c_2=$ \$36 day$^{-1}$ \\
        Border closure cost per unit time & $b=$ \$33.8 million day$^{-1}$ \\
        Border-related outbreak frequency & $r=1/150$ day $^{-1}$ \\
        Multiplicative effect of TTI measures on transmission & $\alpha=0.8$ \\
        Time to detection of border-related outbreaks & $t_\mathrm{det}=14$ days \\
        Proportion of economy restricted to control border-related outbreaks & $p_\mathrm{cont}=0.4$ \\
          \hline
     \end{tabular}
     \caption{Parameter values. See Supplementary Material sec. 3 for further information. All costs are in New Zealand dollars (NZD).}
     \label{tab:params}
 \end{table}

\section{Results}

\begin{figure}
    \centering
    \includegraphics[trim={2cm 0 2cm 0},clip,width=\linewidth]{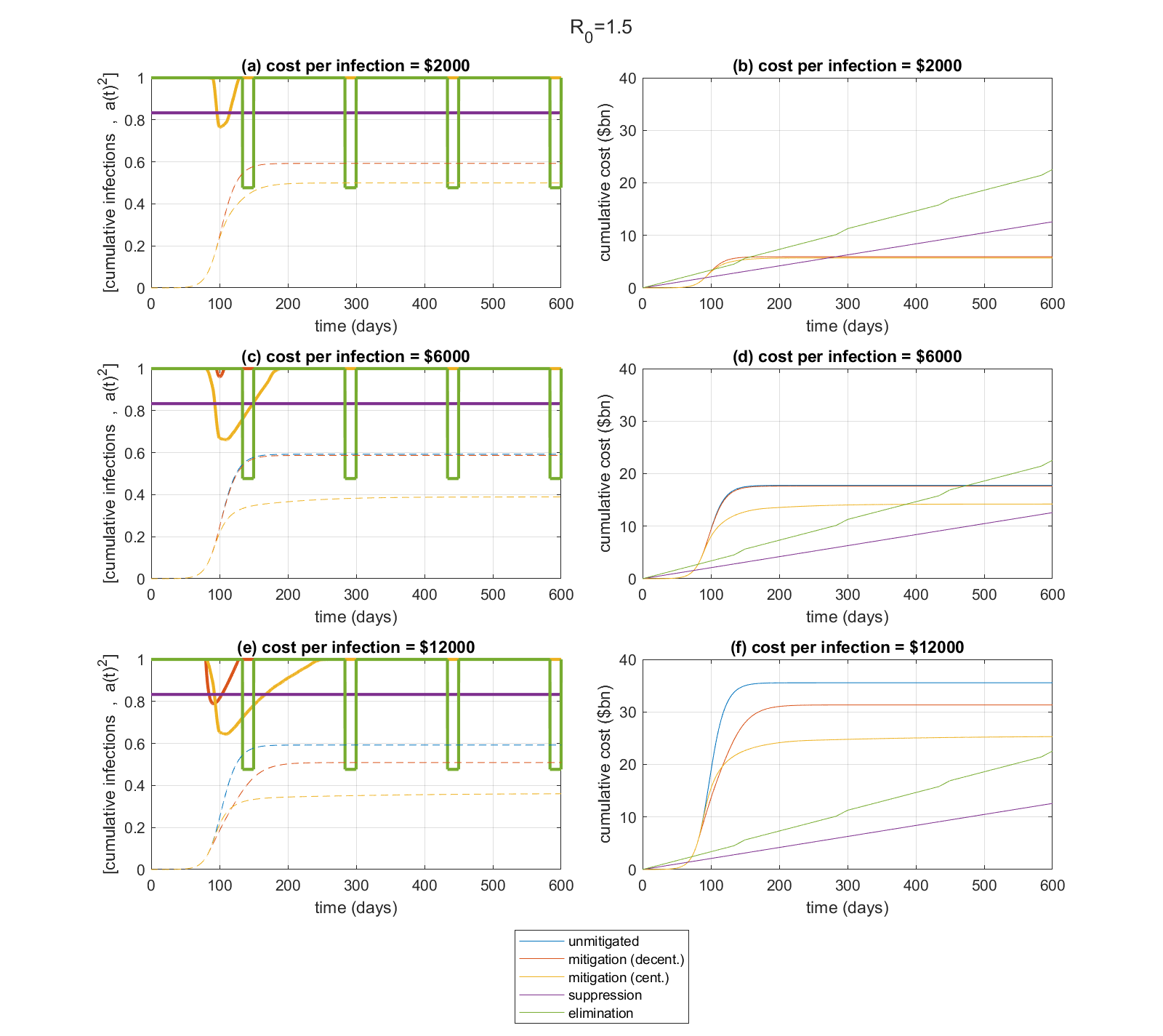}
    \caption{Comparison of the unmitigated (blue), decentralised mitigation (red), centralised mitigation (yellow), suppression (purple), and elimination (green) responses for $R_0=1.5$ and three values of the cost per infection $k$: (a,b) $k=$ \$2000; (c,d) $k=$ \$6000; (e,f) $k=$ \$12000. Left-hand plots (a,c,e) show the time-dependent transmission rate $a(t)^2$ relative to the unmitigated case (solid curves) and the cumulative proportion of the population infected (dashed curves). Note that infections are assumed to be negligible for suppression and elimination. Right-hand plots (b,d,f) show the cumulative cost. Parameter values as in Table 1. }
    \label{fig:1}
\end{figure}

\begin{figure}
    \centering
    \includegraphics[trim={2cm 0 2cm 0},clip,width=\linewidth]{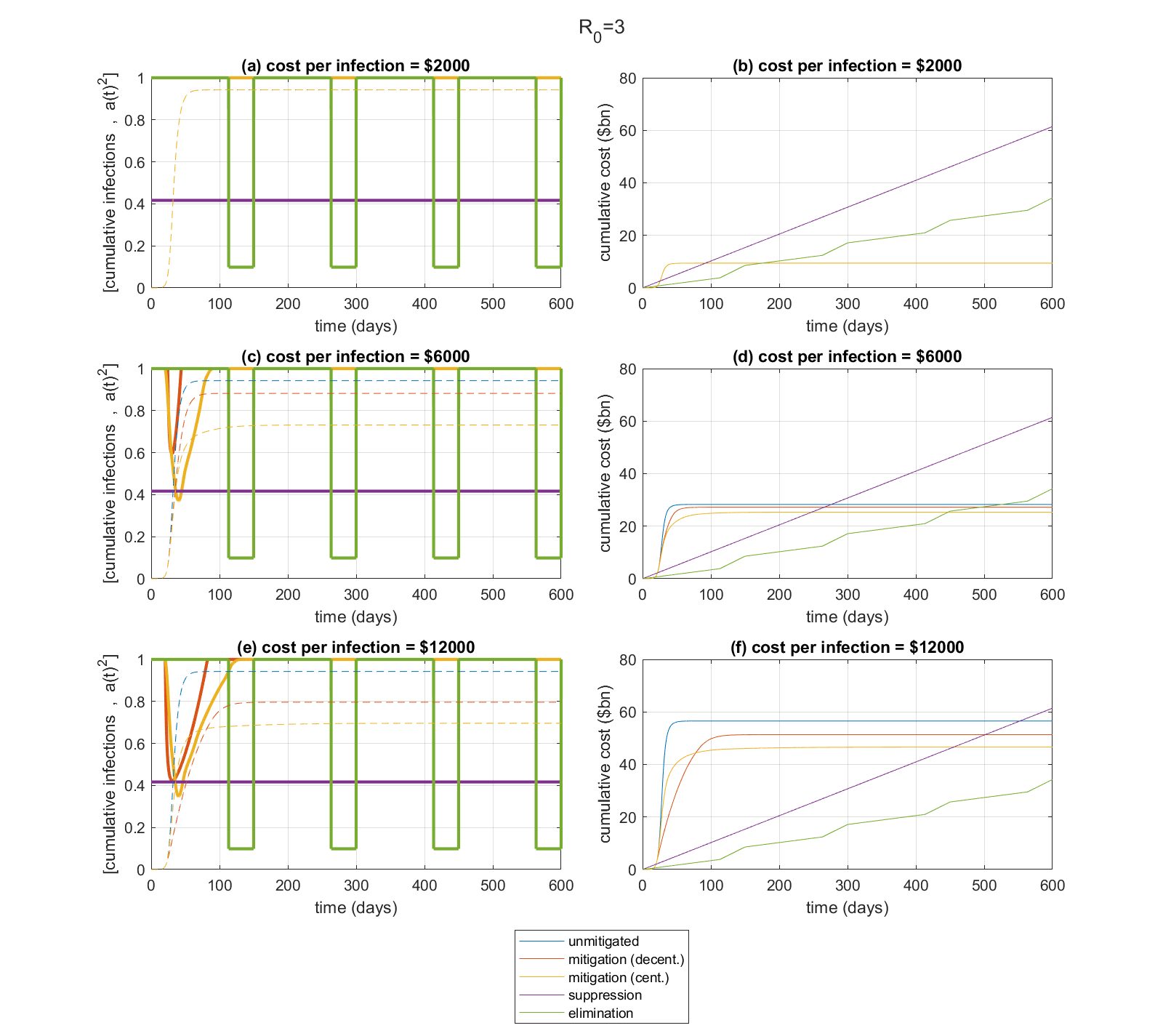}
    \caption{Comparison of the unmitigated (blue), decentralised mitigation (red), centralised mitigation (yellow), suppression (purple), and elimination (green) responses for $R_0=3$ and three values of the cost per infection $k$: (a,b) $k=$ \$2000; (c,d) $k=$ \$6000; (e,f) $k=$ \$12000. Left-hand plots (a,c,e) show the time-dependent transmission rate $a(t)^2$ relative to the unmitigated case (solid curves) and the cumulative proportion of the population infected (dashed curves). Note that infections are assumed to be negligible for suppression and elimination. Right-hand plots   (b,d,f) show the cumulative cost. Parameter values as in Table 1. }
    \label{fig:2}
\end{figure}

Figures \ref{fig:1}-\ref{fig:2} show results for 6 scenarios: low and high transmissibility ($R_0=1.5$ or $R_0=3$) and low, medium and high cost per infection ($k$ = \$2000, \$6000 or \$12000). 
The optimal centralised (yellow curves) and decentralised (red curves) mitigation responses reduce the activity level $a(t)$ during periods of high epidemic activity,  {\hl which reduces the cumulative number of people infected (dashed curves in Figs. \ref{fig:1}-\ref{fig:2}a,c,e). The magnitude of the reduction increases with $R_0$ and with the cost per infection $k$. In all cases, the decentralised response results in a smaller reduction in activity than the centralised response and has a higher aggregate cost.} This is because, in the decentralised response, individuals only consider the trade-off between the cost of reducing activity and the reduction in their own risk, and do not consider the impact of their behaviour on the aggregate epidemic dynamics \cite{gersovitz2004economical,auld2025economics}. Interestingly, when $R_0=3$ and the cost per infection is low ($k=$ \$2000), neither the decentralised nor the centralised mitigation response reduce activity levels at all ($a(t)=1$ for all $t$ in Figure \ref{fig:2}a). This is a consequence of the nonlinear relationship between $R_0$ and final epidemic size \cite{miller2012note}: when $R_0$ is relatively large, small reductions in activity make little difference to the epidemic, and when $k$ is low, larger reductions in activity are not cost-effective. Allowing state-dependent activity levels in the decentralised model produces a stronger response than when the behaviour is independent of disease state, with an aggregate cost that is closer to that of the centralised mitigation model (see Supplementary Figures S4--S5).

The suppression response (Figs. \ref{fig:1}-\ref{fig:2}, purple curves) involves reducing the activity level by a fixed amount to reduce $R_\mathrm{eff}$ to 1. The elimination response (green curves) involves periods of normal activity ($a(t)=1$) and periods during which activity is reduced to eliminate border-related outbreaks (Figure \ref{fig:1}, green). {\hl Suppression has a lower cost than elimination when $R_0=1.5$ (Figure \ref{fig:1}b), but a higher cost when $R_0=3$ (Figure \ref{fig:2}b). This is because, when $R_0$ is small, a modest reduction in activity level combined with TTI measures can reduce $R_\mathrm{eff}$ to 1. In contrast, when $R_0$ is high, reducing $R_\mathrm{eff}$ to 1 is more costly than excluding the pathogen via border controls.}

\begin{figure}
    \centering
    \includegraphics[trim={0.5cm 0 0.5cm 0},clip,width=\linewidth]{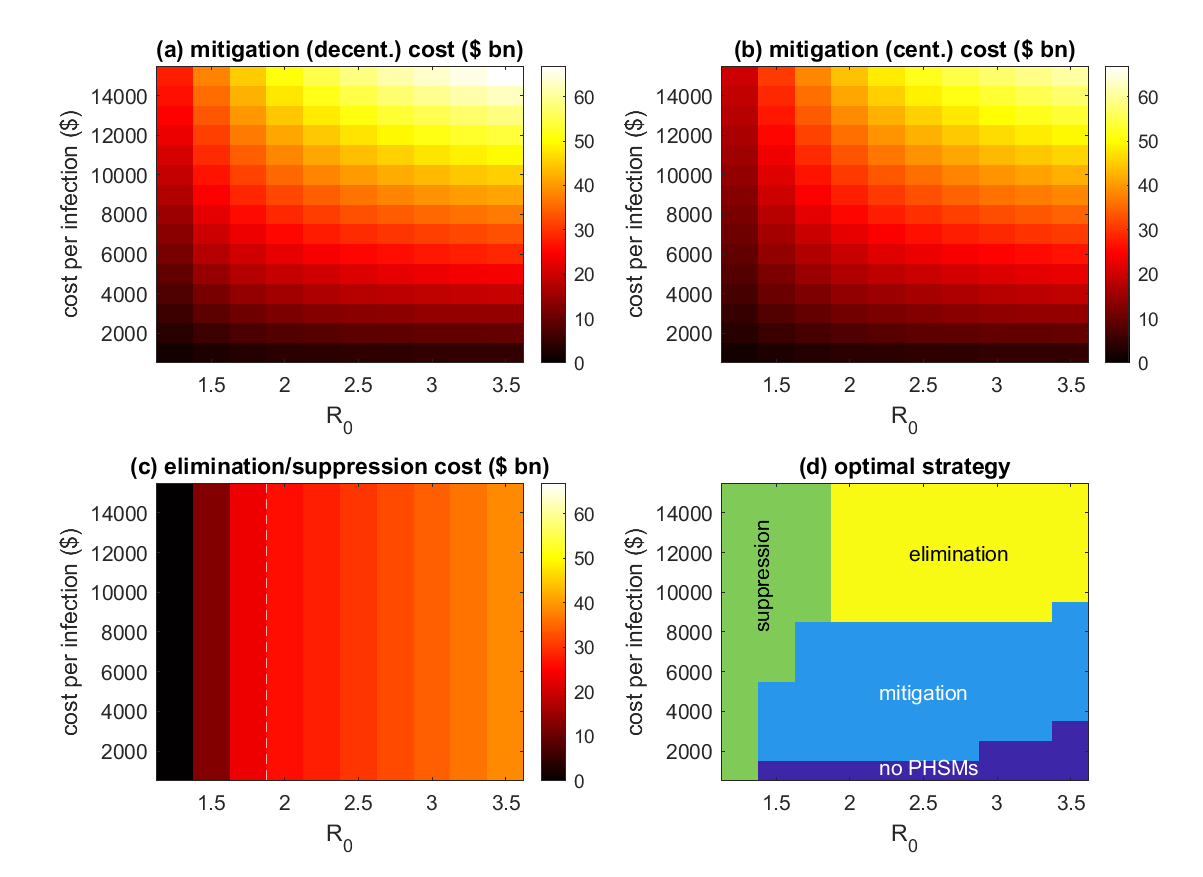}
    \caption{Aggregate cost of different response types at $T=600$ days for a range of values of $R_0$ and cost per infection $k$: (a) a decentralised mitigation response; (b) centralised mitigation response; (c) an elimination or suppression response (whichever has the lower cost). The vertical dashed white line in (c) separates suppression and elimination strategies: suppression is more cost-effective to the left of the dashed line; elimination is more cost-effective to the right of the dashed line. Panel (d) shows which strategy has the lowest cost out of no PHSMs, centralised mitigation, suppression and elimination. In the region labelled `no PHSMs', the centralised mitigation response does not reduce activity levels $a(t)$ by more than $0.1$\% below normal at any time.  }
    \label{fig:3}
\end{figure}

Figure \ref{fig:3} shows the aggregate cost of different response types over a $T=600$ day time horizon for a range of values of $R_0$ and cost per infection $k$. The costs of the decentralised and the centralised mitigation responses increase with both $R_0$ and $k$ (Figure \ref{fig:3}a-b). The cost of elimination or suppression increases with $R_0$ and is independent of $k$ because the number of infections is assumed to be negligible under these strategies (Figure \ref{fig:3}c). Figure \ref{fig:3}d shows the most cost-effective strategy over a $T=600$ day time horizon for different values of $R_0$ and $k$. There is a threshold value of $R_0$ above which elimination is more cost-effective than suppression. {\hl Comparing Eqs. \eqref{eq:sup_cost} and \eqref{eq:elim_cost} shows that this threshold is an increasing function of the cost of border closure $b$ and of the frequency $r$ and time-to-detection $t_\mathrm{det}$ of border-related outbreaks.}

Mitigation tends to be the most cost-effective response when $k$ is relatively low. The threshold value of $k$ above which elimination or suppression is more cost-effective than mitigation increases slightly with $R_0$ and is approximately \$8500 when $R_0>1.5$. Our estimated cost per infection of \$6500 to \$12000 for SARS-CoV-2 in New Zealand in 2020 (see Supplementary Material sec. S3) puts it close to or above the threshold at which elimination is predicted to be more cost-effective than mitigation. 
When the cost per infection $k$ is very low, the centralised mitigation response does not reduce activity level at all (Figure \ref{fig:3}d, dark blue region), meaning population-wide PHSMs are not cost-effective, although targeted control measures such as TTI may still be cost-effective. We also calculated the difference in cost between the decentralised and centralised mitigation strategies (see Supplementary Figure S6). {\hl This shows that the marginal net benefit of centrally imposed PHSMs relative to a no-intervention scenario is small when the cost per infection $k$ is low, meaning that the rationale for intervention is relatively weak.  }

\begin{figure}
    \centering
    \includegraphics[width=\linewidth]{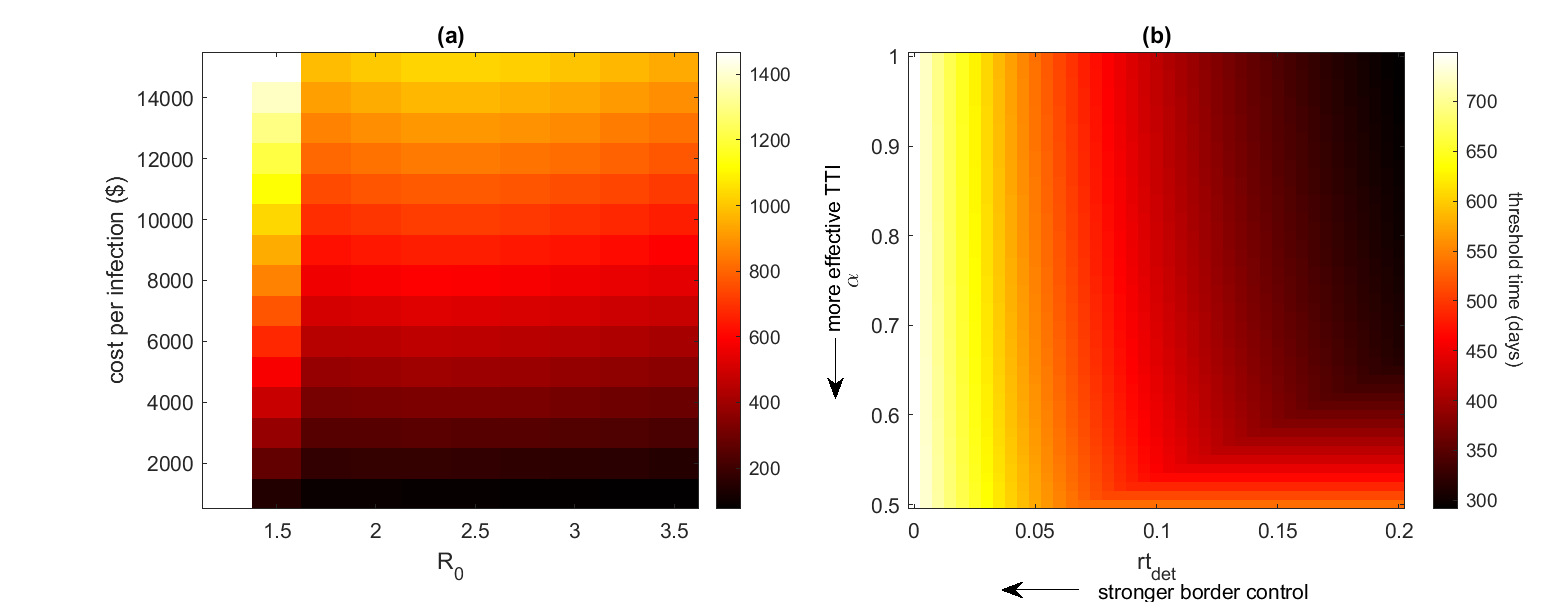}
    \caption{Threshold time $T_\mathrm{crit}$ for vaccine availability below which elimination or suppression becomes more cost-effective than mitigation for a range of values of: (a) $R_0$ and the cost per infection $k$; (b) product of the border-related outbreak frequency and detection time ($rt_\mathrm{det}$) and the TTI transmission multiplier ($\alpha$) for $R_0=3$ and $k=$ \$6000. Other parameter values as in Table \ref{tab:params}. Lighter colours indicate that elimination or suppression is more cost-effective than mitigation over longer time horizons. Note that in the white part of panel (a) where $R_0=1.25$, suppression is always optimal because in the model this can be achieved with TTI measures alone and so does not require any population-level reduction in contact rates.  }
    \label{fig:4}
\end{figure}

{\hl The cumulative average cost of the suppression and elimination responses increases linearly with time (see Eqs. \eqref{eq:sup_cost} and \eqref{eq:elim_cost}), whereas the cumulative cost of the decentralised and centralised mitigation responses increases during the period of epidemic activity before levelling out. Therefore, the most cost-effective strategy will depend on the planner's time horizon.} Figure \ref{fig:4} shows the threshold time horizon $T_\mathrm{crit}$ below which elimination or suppression is more cost-effective than mitigation, for a range of parameter values. 
{\hl If an effective vaccine is expected to become available before $T_\mathrm{crit}$, elimination or suppression is more cost-effective, otherwise mitigation is more cost-effective. The threshold time is an increasing function of the cost per infection $k$. The threshold time is large when $R_0$ is small, but relatively insensitive to $R_0$ for values of $R_0$ above around 1.75 where elimination is preferred to suppression. Fig. \ref{fig:4}b shows that,  for $R_0=3$ and $k=$ \$6000, the threshold time is a decreasing function of the border-related outbreak frequency $r$, outbreak detection time $t_\mathrm{det}$, and TTI transmission multiplier $\alpha$.}

\section{Discussion}

We have developed a mathematical framework for comparing the health and economic costs of different types of pandemic response strategy: mitigation, suppression and elimination. This allows us to determine the most cost-effective strategy in different parts of parameter space. We found that mitigation (i.e. a `flatten the curve' approach) tends to be the most cost-effective strategy when the average cost per infection is low, i.e. for a relatively mild disease. When cost per infection is higher, suppression is the most cost-effective strategy if $R_0$ is relatively low, because the effective reproduction number can be reduced to $1$ with a modest reduction in activity. When cost per infection and $R_0$ are both high, the most cost-effective strategy is elimination, using a combination of border controls to minimise imported infections and PHSMs when necessary to eliminate sporadic outbreaks. 

Our results are not intended to provide a definitive answer to the question of what is the optimal pandemic response strategy, which will depend on a broader range of considerations than those included in our model, but rather to support planning and decision-making for a range of potential pandemic scenarios. Our framework elucidates how the health and economic costs of different strategies depend on epidemiological, economic and public health variables. Key inputs include the basic reproduction number $R_0$, {\hl which is one of the most important determinants of the epidemiological dynamics \cite{heffernan2005perspectives,roberts2007pluses}, and the cost of control measures relative to the cost of infection, which will determine the strength of intervention that is cost-effective \cite{eichenbaum2021macroeconomics}. Disease transmissibility and severity metrics} have been used previously to categorise emerging infectious disease threats \cite{reed2013novel,french2025preparing}. 
Other important factors, which will depend on characteristics of both the pathogen and the surveillance and public health system, include the effectiveness of TTI measures, the frequency of border-related outbreaks, and the speed with which they can be detected. 

Our framework shows how the most cost-effective strategy depends on the expected time $T$ until an effective vaccine or therapy becomes available. 
{\hl Mitigation tends to be more cost-effective when $T$ is larger, because the cost of eliminating or suppressing the diseases increases linearly in $T$.}
A natural extension to the fixed time horizon would be to incorporate the effects of uncertainty in $T$ into the framework (see e.g. \cite{farboodi2021internal}).

We aimed to develop a broadly applicable framework for comparing alternative responses to future pandemic threats, as opposed to a model that faithfully captures all the details of any specific pandemic. Nevertheless, it is useful to contextualise our results in relation to New Zealand's response to Covid-19. The New Zealand Royal Commission of Inquiry into the Covid-19 pandemic found that the elimination strategy was successful at minimising health impacts while also allowing ``the country to spend less time under strict lockdown conditions than many other parts of the world'' \cite{RCOI2024}. Our results are broadly consistent with this finding: in the model, a mitigation strategy had a comparable or higher aggregate cost than elimination under parameter values estimated from New Zealand data. Our elimination model anticipates an average of four periods of stringent PHSMs lasting around one month each (see Figure \ref{fig:2}), applying to 40\% of the economy. It does not allow for a longer period of wider-scale PHSMs required to control a larger initial outbreak: between March and June 2020, New Zealand spent 49 days with strict stay-at-home orders nationally, and a further 26 days with less stringent measures before most community controls were removed on 8 June 2020 \cite{baker2020successful}. Also, our model does not anticipate the emergence of variants that are more transmissible ($R_0$) and/or more severe (higher $k$), which impacted the dynamics of the pandemic and the costs of different responses globally. On the other hand, the mitigation model does not include costs arising from an overwhelmed healthcare system and associated societal harms and may therefore underestimate costs in scenarios where this occurs.

Our framework highlights the role of modifiable public health factors in reducing the costs of a future pandemic response. For example, a well-resourced testing and contact tracing system and support for isolation measures (reducing $\alpha$) can help alleviate the need for blanket PHSMs. {\hl Robust border control systems and quarantine facilities can lower the risk of border-related outbreaks (reducing $r$) while effective surveillance systems can promote early detection (reducing $t_\mathrm{det}$).} This could include wastewater testing, which was successfully used as an early warning system during the Covid-19 pandemic \cite{hewitt2022sensitivity} and for other viral pathogens \cite{keshaviah2023wastewater}. {\hl Our results show that these factors can lower the expected cost of an elimination response.} Characteristics of the pathogen also play a role. Diseases for which presymptomatic or asymptomatic transmission is rare (e.g. {\hl severe acute respiratory syndrome (SARS)}) are generally more amenable to TTI and quarantine measures, since symptomatic individuals are easier to identify and isolate \cite{thompson2026infectious}. {\hl An elimination response in these cases would therefore be less costly than for comparable diseases with high rates of asymptomatic transmission (e.g. influenza, SARS-CoV-2). }

A strength of our model is that it depends on a relatively small number of parameters, most of which would be routinely estimated for a novel pathogen in the early stages of a pandemic threat. Therefore, our model could be rapidly deployed with disease-specific parameter ranges to support strategy selection in the early stages of a future pandemic. Another strength is that our model explicitly quantifies the endogenous behavioural response that would be expected to an infectious disease threat in the absence of centrally imposed PHSMs. This allows the marginal benefit of introducing control measures to be estimated, providing a more nuanced rationale for interventions. 

We estimated costs of border restrictions and frequency of border-related outbreaks from the experience of Covid-19 in New Zealand in 2020-21. In this period, all international arrivals to New Zealand were required to spend 14 days in government-managed quarantine facilities. An alternative strategy would be to use looser border controls (e.g. home quarantine) and accept the higher likelihood of border-related outbreaks. These options could be explored in our framework with a lower border control cost $b$ and higher outbreak frequency $r$.   
The efficacy and feasibility of the strategies we have investigated will differ between jurisdictions \cite{martignoni2024sars}. New Zealand's relatively remote geographical location and lack of land borders made reducing imported infections easier than in other countries. Jurisdictions with higher volumes of passenger travel and land-based trade would likely have a higher frequency of border-related outbreaks ($r$) and potentially also higher cost of border restrictions ($b$). Our model shows that these factors mean an elimination strategy is less likely to be cost-effective. Furthermore, if there is widespread undetected transmission before control measures are introduced, such as occurred with SARS-CoV-2 in the {\hl United States and United Kingdom} in early 2020 \cite{bedford2020cryptic,duplessis2021establishment}, then eliminating the initial outbreak is likely to be infeasible or so costly that it is not desirable.

Our study has several important limitations. The models we have used for transmission dynamics, behavioural response, and costs are, by design, highly simplified. This enables the dependence of outcomes on {\hl some of the key epidemiological and economic parameters to be systematically explored. However, other aspects that are not included in the model also may have a significant influence, so} our results should not be interpreted as precise predictions or recommendations. 

The endogenous behavioural response model assumes that individuals are self-interested rational actors, with perfect knowledge of the epidemic dynamics and the cost functions. In reality, people may underestimate or overestimate the risk posed by the disease. The cost of reducing activity levels is not the same for everyone, and people with low income, essential workers, the elderly, children/students, ethnic minorities, and people with disabilities are just some of the population groups identified as having a high cost of reducing activity levels \cite{li2023scoping,garnier2021socioecon}. People may behave with some degree of altruism \cite{farboodi2021internal}, particularly in relation to family members, friends and in small communities where most interactions are with other locals \cite{lehmann2006evolution,debarre2012evolution}. People may also behave differently when they believe they are currently or previously infected, compared with when they believe they are susceptible \cite{fenichel2013economic}. Future work will incorporate some of these aspects into the model framework. 

Our model assumes that the pandemic pathogen continues to transmit globally, and considers the decision by an individual jurisdiction between adopting a mitigation, suppression or elimination strategy. This does not allow for the possibility that the pathogen could be controlled globally, as occurred for example with the SARS outbreaks in 2003 \cite{heymann2004international}. This may be a feasible outcome, particularly for a pathogen with modest $R_0$, relatively high severity, and low rates of asymptomatic transmission. However, a quantitative analysis of this is out of scope of the current study.

Some types of costs, risks and considerations are not included in our model and would need to be factored into decision making. Examples include: the  social and psychological impacts of prolonged periods of stringent PHSMs (e.g. impacts on education, childhood development, family separation, and mental health) \cite{zachreson2024ethical}; the need to maintain social license for control measures \cite{RCOI2024}; the feasibility of strategic aims \cite{martignoni2024sars}; the potential consequences of pathogen evolution \cite{antia2003role,davies2021estimated}; and the impact of high infection rates on healthcare systems and health workers \cite{fong2024nhs}. A risk of the elimination and suppression strategies is that, at some point in time, these become infeasible to maintain \cite{metcalf2021challenges}, for example because border-related outbreaks become too frequent, due to emergence of a more transmissible variant, due to {\hl pandemic fatigue} or the erosion of the social license for PHSMs, or because vaccines are slower or less effective than anticipated. This could force a switch to a mitigation strategy while the population is still fully susceptible, potentially meaning that all the costs of a mitigation strategy are incurred in addition to some of the costs of elimination or suppression. 
On the other hand, switching strategy from elimination or suppression to mitigation is generally possible, whereas switching the other way is usually impossible. {\hl Elimination strategies in particular require a sustained high level of compliance with control measures, which relies on clear communication of their rationale and high levels of social buy-in and trust.}
Decision makers need to weigh a range of factors in determining the best strategy and our framework is intended to help provide one source of evidence, not to give definitive recommendations.

The model ignores important heterogeneities in transmission rates, disease severity and economic costs between different groups, and only considers aggregate utility across the whole population. It is also important to consider how the costs both of the epidemic itself and of any interventions that are used are distributed among different parts of the population \cite{silva2025ethical}, including different socioeconomic and ethnicity groups \cite{bedson2021review,zelner2022there}. Future work will address this by applying the cost-benefit framework developed in this study to stratified models (e.g. \cite{ma2021modeling,harris2025population,datta2025modelling}) that capture some of the heterogeneity in transmission rates, cost of infection, and cost of reducing contact rates.

\subsection*{Author contributions}
Conceptualization: MJP, AH, SCH.
Formal analysis: MJP, MS, TFT, SCH.
Investigation: MJP, MS, TFT, SCH.
Methodology: MJP, SCH.
Software: MJP.
Supervision: MJP.
Visualization: MJP.
Writing -- original draft: MJP, SCH.
Writing -- review and editing: all.

\subsection*{Data availability statement}
Code to reproduce the results in this article is available at \cite{github-repo}.

\subsection*{Conflict of interest statement}
The authors have no conflicts of interest to declare.

\subsection*{Funding statement}
MJP was supported by the Marsden Fund grant (24-UOC-020) managed by the Royal Society Te Ap\=arangi. MJP, MS and TFT were supported by the project ``Improving models for pandemic preparedness and response: modelling differences in infectious disease dynamics and impact by ethnicity'' (TN/P/24/UoC/MP) funded by Te Niwha, the Infectious Diseases Research Platform, co-hosted by PHF Science and the University of Otago and provisioned by the Ministry of Business, Innovation and Employment, New Zealand. AH was supported by a Natural Sciences and Engineering Research Council of Canada Discovery Grant (RGPIN RGPIN-2023-05905). AH and MJP are part of a Collaborative Research Group award from the Atlantic Association for Research in the Mathematical Sciences. RNT acknowledges the support of the JUNIPER partnership (grant number MR/X018598/1).

\subsection*{Acknowledgements}
 The authors are grateful to Tim Ng and two anonymous reviewers for helpful feedback on an earlier version of this manuscript.

 \bibliography{references}

\end{document}


\maketitle


\section{Supplementary methods}

\subsection{Centralised mitigation response} \label{sec:centralised}

The time-varying activity level $a(t)$ in Eq. (3) of the main text was approximated by a piecewise cubic Hermite interpolating polynomial with knots at equally spaced time points $t_j= j\tau$ where the value of the control function is specified by $a(t_j)=a_j$ ($j=0,1,\ldots, T/\tau$). We used a time spacing between knots of $\tau=10$ days. This approximation resulted in a finite-dimensional optimization problem for ${\bf a}=[a_0, \ldots, a_n]$, which we solved using the {\em fmincon} routine in Matlab, which uses an `interior-point' algorithm.

\begin{figure}
    \centering
    \includegraphics[trim={3cm 3cm 3cm 3cm},clip,width=.9\textwidth]{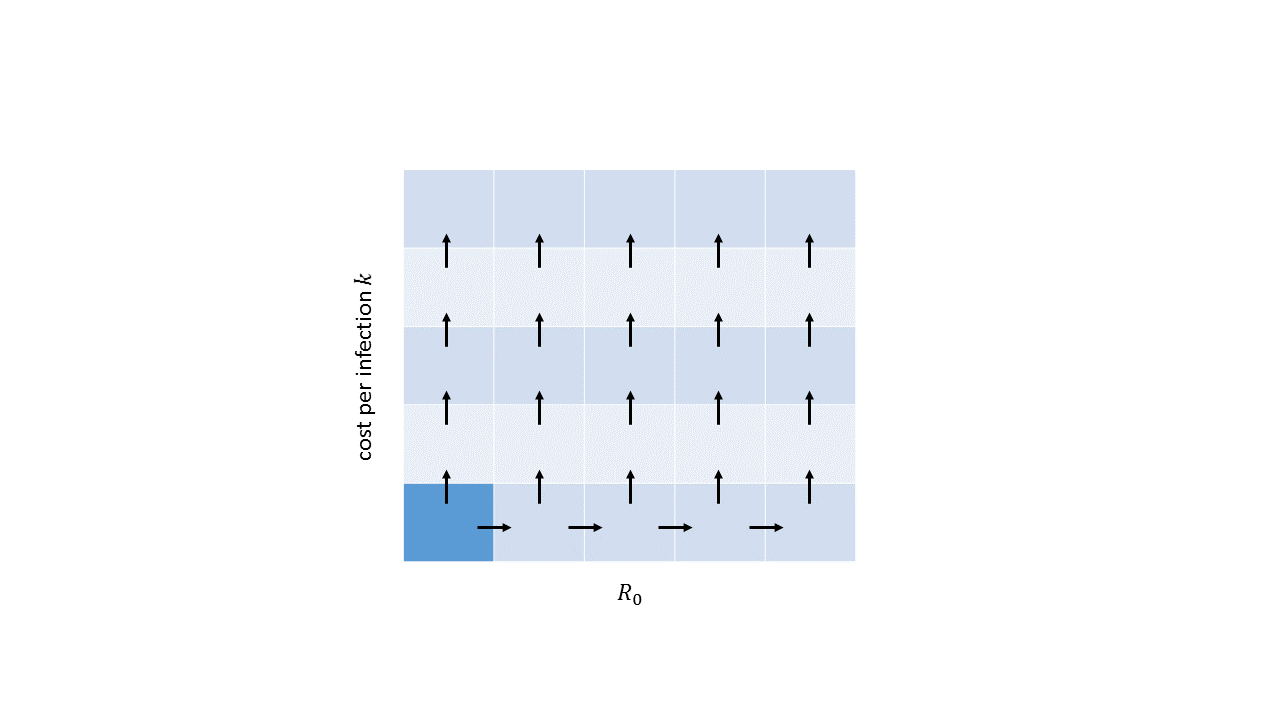}
    \caption{Schematic diagram showing how the optimization algorithm was initialised at each combination of parameter values for $R_0$ and cost per infection $k$. The first parameter combination (dark blue square) was initialised at the unmitigated state ($a(t)=1$ for all $t\ge 0$). All other parameter combinations were initialised at the solution from a previous parameter combination, as indicated by the black arrows. }
    \label{fig:init_diagram}
\end{figure}

The optimization problem was solved for a range of values of the parameters $R_0$ and $k$. To ensure convergence, the optimization routine needs to be initialised reasonably close to the minimum of the objective function. We started with the smallest value of $R_0$ and the smallest value of $k$ (where we expect any reduction in contact rates to be relatively small) and initialised the optimization routine in the unmitigated state ($a(t)=1$ for all $t\ge 0$). For each subsequent combination of $R_0$ and $k$, the optimization routine was initialised with the optimal solution from a previous combination of $R_0$ and $k$ (see Figure \ref{fig:init_diagram}). 

\subsection{Decentralised mitigation response}

Consider the situation where the population has activity level $a(t)$, and let the functional $C\left[\tilde{a} | a\right]$ denote the cost to a single individual $i$ who chooses a different activity level $\tilde{a}(t)$. We assume that the population size $N$ is sufficiently large that the aggregate epidemic dynamics are insensitive to the behaviour of a single individual, so that $S(t)$ and $I(t)$ solve Eqs. (1)--(2). In this section, we consider the simplest version of the problem in which individuals do not modify their behaviour according to which disease compartment they are in (see Supplementary Material sec. 1.4 for a more general, state-dependent version of the model). Then the functional $C\left[\tilde{a}|a\right]$ is given by
\begin{equation} \label{eq:individual_cost}
    C\left[\tilde{a}|a\right] = k\left(1-S_{i,\infty}\left[\tilde{a}|a\right] \right) + c_1\int_0^\infty \left(1-\tilde{a}(t)\right) dt + c_2\int_0^\infty \left(1-\tilde{a}(t)\right)^2 dt
\end{equation}
where $S_{i,\infty}\left[\tilde{a}|a\right]= \lim_{t\to\infty} S_i(t)$ is the probability that individual $i$ remains uninfected throughout the epidemic, which is governed by
\begin{equation}
    \frac{dS_i}{dt} = -\beta \tilde{a}(t) a(t) S_i \frac{I}{N} \label{eq:dSidt_nash} 
\end{equation}
where $I(t)$ solves Eqs. (1)--(2). This equation encodes the fact that the force of infection on individual $i$ is controlled by the product of individual $i$'s activity level $\tilde{a}(t)$, the population activity level $a(t)$, and the infection prevalence $I(t)/N$. 
A Nash equilibrium is a function $a^*(t)$ such that 
\begin{equation}
    C\left[a^*|a^*\right] \le C\left[\tilde{a}|a^*\right]
\end{equation}
for all functions $\tilde{a}(t)$ \cite{reluga2010game}. 

Because the aggregate epidemic dynamics are independent of $\tilde{a}(t)$, this problem is amenable to analysis. It follows from Eq. \ref{eq:dSidt_nash} that the probability of individual $i$ avoiding infection $S_{i,\infty}\left[\tilde{a}| a\right]$ is given by
\begin{equation}
   S_{i,\infty}\left[\tilde{a}|a \right] = \exp\left(-\frac{\beta}{N}\int_0^\infty \tilde{a}(t) a(t) I(t) dt \right)
\end{equation}
Substituting this expression into Eq. \eqref{eq:individual_cost} and differentiating the functional with respect to $\tilde{a}(t)$ gives
\begin{equation}  \label{eq:cost_deriv}
\frac{\delta C}{\delta \tilde{a}} = \frac{k\beta}{N} S_{i,\infty}  a(t)I(t) -c_1  - 2c_2\left(1-\tilde{a}(t)\right)
\end{equation}
A Nash equilibrium requires that $\tilde{a}(t)=a(t)$ and either: (i) $\delta C/\delta \tilde{a}(t)=0$; (ii) $\delta C/\delta \tilde{a}(t)<0$ and $\tilde{a}(t)=1$; or (iii) $\delta C/\delta \tilde{a}(t)>0$ and $\tilde{a}(t)=0$. 
Evaluating Eq. \eqref{eq:cost_deriv} at $a(t)=\tilde{a}(t)$ gives
\begin{equation} \label{eq:deriv_eval}
 \left.\frac{\delta C}{\delta \tilde{a}}\right|_{a(t)=\tilde{a}(t)}  = \left(\frac{k\beta}{N} S_{i,\infty} I(t) + 2c_2 \right)\tilde{a}(t)- (c_1+2c_2)
\end{equation}
Noting that a change in $\tilde{a}(t)$ for a single value of $t$ has a negligible effect on $S_{i,\infty}$, Eq. \eqref{eq:deriv_eval} is of the form $-\alpha_0 + \alpha_1 \tilde{a}(t)$ where, for given $t$, $\alpha_0$ and $\alpha_1$ are positive numbers that are independent of $\tilde{a}(t)$. Hence $\left.\delta C/\delta \tilde{a}\right|_{a(t)=\tilde{a}(t)}$ is negative when $0\le a(t)<\alpha_0/\alpha_1$, meaning that if the population activity level is $a(t)$, an individual can reduce their cost by increasing their activity level. The converse is true when $a(t)>\alpha_0/\alpha_1$. Therefore, if $\alpha_0/\alpha_1<1$, there is a unique solution for the Nash equilibrium at $a^*(t)=\alpha_0/\alpha_1$. If $\alpha_0/\alpha_1>1$ then the Nash equilibrium is given by $a^*(t)=1$ since any individual reducing their activity level below $1$ will experience a higher cost. It follows that
\begin{equation} \label{eq:equilib}
        a^*(t) = \mathrm{min}\left[1, \frac{c_1+2c_2}{k\beta S_{i,\infty}^* I^*(t)/N + 2c_2}\right]
\end{equation}
This equation defines $a^*(t)$ in terms of the time-dependent infection prevalence $I^*(t)$ and the probability of avoiding infection $S_{i,\infty}^*$ when the whole population has activity level $a^*(t)$. The values of $I^*(t)$ and $S_{i,\infty}^*$ are defined by the solutions of Eqs. (1), (2) and \eqref{eq:dSidt_nash} with $\tilde{a}(t)=a(t)=a^*(t)$. 

We calculated the solution to Eq. \eqref{eq:equilib} using the natural fixed-point iteration scheme. This entailed: (i) choosing an initial $a^{(n)}(t)$ for $n=0$; (ii) solving Eqs. (1), (2) and \eqref{eq:dSidt_nash} with $\tilde{a}(t)=a(t)=a^{(n)}(t)$ to obtain $I^{(n)}(t)$ and $S_i^{(n)}(t)$; and (iii) iterating $a^{(n)}(t)$ according to
\begin{equation}
    a^{(n+1)}(t) = \mathrm{min}\left[ 1, (1-\omega_n)a^{(n)}(t) + \omega_n \frac{c_1+2c_2}{k\beta S_{i,\infty}^{(n)} I^{(n)}(t)/N + 2c_2 }\right], \qquad n=0,1,\ldots
\end{equation}
where $\omega_n\in[0,1]$ is a relaxation factor, which may be chosen to promote convergence. We set $\omega_n=0.3+0.7e^{-0.0015n}$, which progressively reduced the magnitude of the change in successive iterations of $a^{(n)}(t)$ as the iteration count $n$ increased. All calculations were performed by discretising time using the same fixed time step, $\delta t=0.1$ days, as for the ordinary differential equation solution (see Sec. \ref{sec:ODEs}). We continued iterations on $a^{(n)}(t)$ until the following convergence criterion was met:
\begin{equation}
\frac{ \left\| a^{(n+1)}(t) -a^{(n)}(t) \right\|_\infty}{ \left\| a^{(n)}(t) \right\|_\infty} < \varepsilon
\end{equation}
with a tolerance of $\varepsilon=10^{-8}$. 

Similar to the method for the central planner's problem, for the first values of the parameters $R_0$ and $k$, we initialised the fixed-point iteration algorithm at the unmitigated state ($a^{(0)}(t)=1$ for all $t\ge 0$. For subsequent parameter combinations, we initialised $a^{(0)}(t)$ at the converged solution for a previous parameter combination (see Figure \ref{fig:init_diagram}). 

\begin{figure}
    \centering
    \includegraphics[trim={2cm 0 2cm 0},clip,width=\linewidth]{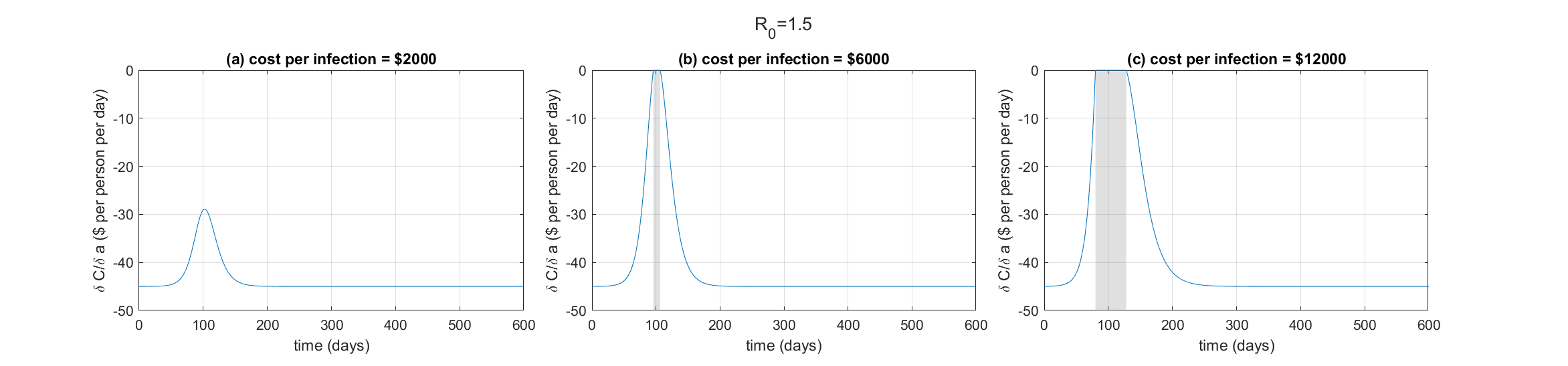}
    \caption{Derivative of the cost functional $\delta C/\delta \tilde{a}$ against time at the equilibrium solution for the decentralised response for $R_0=1.5$ and three values of the cost per infection: (a) $k=$ \$2000; (b) $k=$ \$6000; (c) $k=$ \$12000. Other parameters as shown in Table 1 of the main article. Grey shaded region shows values of $t$ for which $a(t)<1$; elsewhere $a(t)=1$. The plots confirm that either $\delta C/\delta \tilde{a}=0$, or $\delta C/\delta \tilde{a}<0$ and $\tilde{a}(t)=1$ for all values of $t$.  }
    \label{fig:check1}
\end{figure}

\begin{figure}
    \centering
    \includegraphics[trim={2cm 0 2cm 0},clip,width=\linewidth]{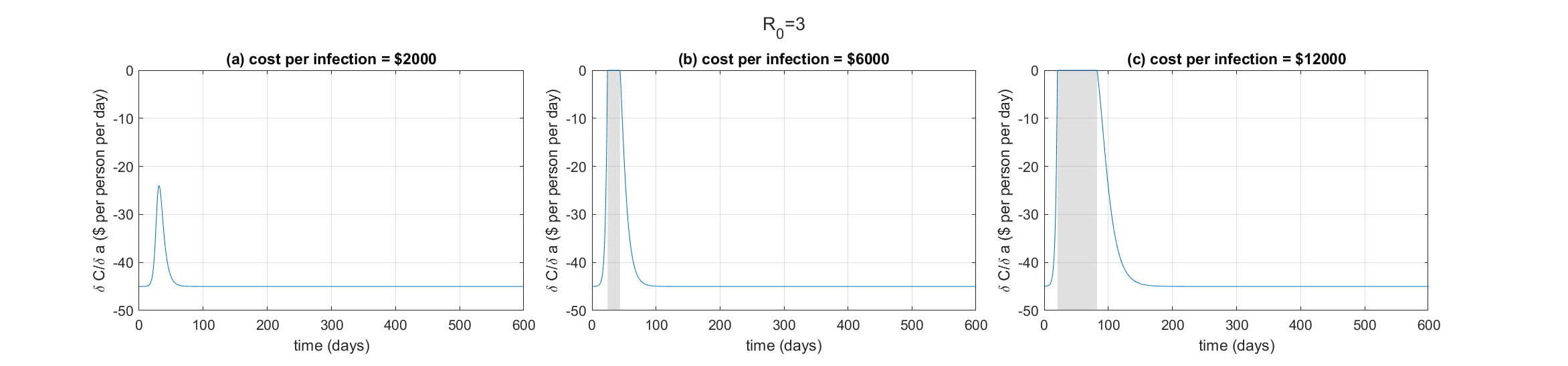}
    \caption{Derivative of the cost functional $\delta C/\delta \tilde{a}$ against time at the equilibrium solution for the decentralised response for $R_0=3$ and three values of the cost per infection: (a) $k=$ \$2000; (b) $k=$ \$6000; (c) $k=$ \$12000. Other parameters as shown in Table 1 of the main article. Grey shaded region shows values of $t$ for which $a(t)<1$; elsewhere $a(t)=1$. The plots confirm that either $\delta C/\delta \tilde{a}=0$, or $\delta C/\delta \tilde{a}<0$ and $\tilde{a}(t)=1$ for all values of $t$.  }
    \label{fig:check2}
\end{figure}

To check our method converged to the correct equilibrium, we plotted $\delta C/\delta \tilde{a}$ against time (Supplementary Figures \ref{fig:check1}--\ref{fig:check2}). This showed that, for all $t\ge 0$, either: (i) $\delta C/\delta \tilde{a}=0$ (meaning $a^*(t)$ is a local minimum of the individual cost functional); or (ii) $\delta C/\delta \tilde{a}<0$ and $a^*(t)=1$ (meaning $a^*(t)$ is at its maximum allowed level and any decrease in $a^*(t)$ incurs a higher cost). This confirms that any individual changing their activity level from the equilibrium value will experience a higher cost. 

\subsection{Ordinary differential equation solution} \label{sec:ODEs}

We solved the system of ordinary differential equations (1), (2) and (5) in the main text numerically with a fixed time step of $\delta t=0.1$ days, assuming a constant force of infection at each time step. The solution at the subsequent time step was calculated according to:
\begin{align}
    S_i(t+\delta t) &= S_i(t) \exp\left( -\lambda_i(t) \delta t \right) \\
    S(t+\delta t) &= S(t) \exp\left( -\lambda(t) \delta t \right) \\
    I(t+\delta t) &= I(t)\exp\left(-\gamma \delta t\right) + S(t)\left[1- \exp\left( -\lambda(t) \delta t \right)\right] 
\end{align}
where the force of infection terms were defined by $\lambda_i(t) = \beta \tilde{a}(t)a(t)I(t)/N$ and $\lambda(t) = \beta a(t)^2 I(t)/N$, where $N$ is the population size.
This method preserves positivity of solutions and was found to be more numerically stable than adaptive step size algorithms when combined with the numerical optimzation routine for $a_j=a(t_j)$ described in Sec. \ref{sec:centralised}. 

We assumed that, at $t=0$, a small fraction $\iota_0=10^{-5}$ of the population was infectious, and the remainder was susceptible, which gives the following initial conditions:
\begin{equation}
S_i(0)=1-\iota_0, \qquad S(0)=(1-\iota_0)N, \qquad I(0)=\iota_0 N,
\end{equation}
The model was run until $T=1100$ days, by which time the epidemic had finished (i.e. $I(t)$ had returned to a negligibly small value) in all cases.

\subsection{Decentralised mitigation model with state-dependent activity levels}

In the main analysis of the decentralised mitigation model, we assumed that individuals' activity levels do not depend on their disease status (susceptible, infectious or recovered). Here, we extend this analysis to include state-dependent activity levels $a_S(t)$, $a_I(t)$ and $a_R(t)$. Note that, although the centralised mitigation model could also incorporate state-dependent activity levels in principle, there are likely to be significant obstacles and limitations on the ability of a central planner to differentially control people depending on their disease state in practice. Therefore, we do not consider this here.

Before formulating the optimisation problem, we make some observations and simplifications. Recovered individuals have no incentive to reduce activity and hence we set $a_R(t)=1$. Infectious individuals also do not have any incentive to reduce activity, because there is no further risk to them from social contacts. Hence, for strictly self-interested individuals with perfect information, we would have $a_I(t)=1$. However, this is not realistic for several reasons: 
\begin{enumerate}
\item Many potential pandemic pathogens exhibit a substantial level of pre-symptomatic and/or asymptomatic transmission. Therefore, even with perfect diagnostic testing, a certain proportion of transmission will occur without people knowing they are infectious, and hence before they are able to switch their activity level from $a_S(t)$ to $a_I(t)$. 
\item For symptomatic infectious individuals, there will be some intrinsic reduction in activity level because some people will be too sick to participate in some types of activity, or will be discouraged from non-essential social interactions by voluntary public health guidance and social norms.
\item People are more likely to reduce activity levels altruistically when they know they pose an active transmission risk to others and can reduce that risk by paying time-limited costs, than they are in a situation where they have no knowledge of their disease status. In the latter case, they would need to make costly reductions in activity over a prolonged period of time, with a low probability of being infectious at any given point in time. 
\end{enumerate}
Point 3 is more speculative than points 1 and 2. Nevertheless, these effects all act in the direction of making infectious individuals reduce activity level, so we make the simplifying assumption that people continue to adopt activity level $a_S(t)$ for the duration of their infectious period, and only return to normal activity levels after the end of their infectious period.

In the likely event that there are limitations on the accuracy, availability, timeliness and uptake of diagnostic testing, there may be some misclassification with respect to the recovered state. There may be some individuals who are: (i) recovered but believe they are susceptible, because they had a mild or asymptomatic infection; and (ii) susceptible but believe they are recovered, because they experienced symptoms consistent with the pathogen but caused by something else. For simplicity, we ignore these biases and assume that recovered individuals have perfect knowledge of their disease state. However, we note that the above effects would mean that: (i) some recovered individuals continue to reduce activity level and therefore incur additional costs; and (ii) activity levels of susceptibles are not reduced as much as is optimal. Therefore, this model is likely optimistic as to the cost-effectiveness with which the epidemic can be controlled, while the state-independent model is likely pessimistic as it assumes that recovered individuals must continue to behave as if they were susceptible.

Under these assumptions, the disease dynamics are governed by the same set of differential equations (Eqs. (1), (2) and (5)) as the state-independent model, because susceptible and infectious individuals both have activity level $a_S(t)$, and recovered individuals do not affect the dynamics. However, the cost functional $C\left[\tilde{a},a\right]$ is different because individuals only pay the cost of reduced activity when in the susceptible or the infectious state:
\begin{equation}
    C\left[\tilde{a},a\right] = k\left(1-S_{i,\infty}\right) + \int_0^\infty \left(S_i(t')+I_i(t')\right)\left( c_1\left(1-\tilde{a}(t')\right) + c_2\left(1-\tilde{a}(t')\right)^2\right) dt'
\end{equation}

Taking the derivative of the cost functional with respect to $\tilde{a}(t)$:
\begin{align}
    \frac{\delta C}{\delta \tilde{a}(t)} &= \frac{k\beta}{N} a(t)I(t)S_{i,\infty} - \left( S_i(t)+I_i(t)\right)\left(c_1 + 2c_2\left(1-\tilde{a}(t)\right)\right)  \nonumber \\
    & +\int_t^\infty \frac{\delta}{\delta \tilde{a}(t)}\left(S_i(t')+I_i(t')\right)\left( c_1\left(1-\tilde{a}(t')\right) + c_2\left(1-\tilde{a}(t')\right)^2\right) dt'
\end{align}

Here, the first term represents the marginal cost due to risk of infection associated with a change in activity level $\tilde{a}(t)$ at time $t$. The second term represents the marginal cost of activity level reduction at time $t$. The third term, consisting of the integral, represents the marginal change in the aggregate cost of future activity level reductions at times $t'>t$ caused by a change in $\tilde{a}(t)$. The first two terms have analogs in the state-independent model (see Eq. (8) of the main text). The third term does not appear in the state-independent model, but appears here because a change in $\tilde{a}(t)$ affects the probability of being in the susceptible or infectious state, and therefore the probability of adopting activity level reductions, at future times $t'>t$.

Noting that 
\begin{equation}
S_i(t)=\exp\left(-\frac{\beta}{N}\int_0^t \tilde{a}(t') a(t') I(t') dt'\right)
\end{equation}
and making the pseudo-steady state approximation $I_i(t)\approx \beta/(\gamma N) S_i(t) I(t)$ leads to
\begin{equation}
    \frac{\delta}{\delta \tilde{a}(t)}\left( S_i(t')+I_i(t') \right) = -\frac{\beta}{N} a(t)I(t)\left(S_i(t')+I_i(t')\right)
\end{equation}

Hence we obtain
\begin{equation}
    \frac{\delta C}{\delta \tilde{a}(t)} = \frac{k\beta}{N} a(t)I(t)S_{i,\infty} - \left( S_i(t)+I_i(t)\right)\left(c_1 + 2c_2\left(1-\tilde{a}(t)\right)\right) -\frac{\beta}{N} a(t)I(t)Q_{\tilde{a}}(t)
\end{equation}
where
\begin{equation}
Q_{\tilde{a}}(t) = \int_t^\infty \left( S_i(t')+I_i(t')\right)\left( c_1\left(1-\tilde{a}(t')\right) + c_2\left(1-\tilde{a}(t')\right)^2 \right) dt'
\end{equation}

Finally, setting $\tilde{a}(t)=a(t)$ and $\delta C/\delta \tilde{a}(t)=0$ gives
\begin{equation} \label{eq:a_state_dependent}
a(t) = \frac{(c_1+2c_2)\left(S_i(t)+I_i(t)\right)}{\frac{k\beta}{N}I(t)S_{i,\infty} + 2c_2\left(S_i(t)+I_i(t)\right) - \frac{\beta}{N}I(t)Q_a(t) }
\end{equation}

Unlike the state-independent model, Eq. \eqref{eq:a_state_dependent} does not provide a closed-form expression for $a(t)$ in terms of the ODE model solution variables, because the expression for $a(t)$ depends on the value of $a(t')$ for $t'>t$ via $Q_a(t)$. Nonetheless, Eq. \eqref{eq:a_state_dependent} can be evaluated by working backwards in time from $t=T$ to $t=0$ and updating $Q_a(t)$ at each step once $a(t')$ for $t'>t$ is known. 
As for the state-independent model, we solve Eq. \eqref{eq:a_state_dependent} using fixed-point iteration, solving the ODE model and updating $a(t)$ at each iteration.

\begin{figure}
    \centering
    \includegraphics[trim={2cm 0 2cm 0},clip,width=\linewidth]{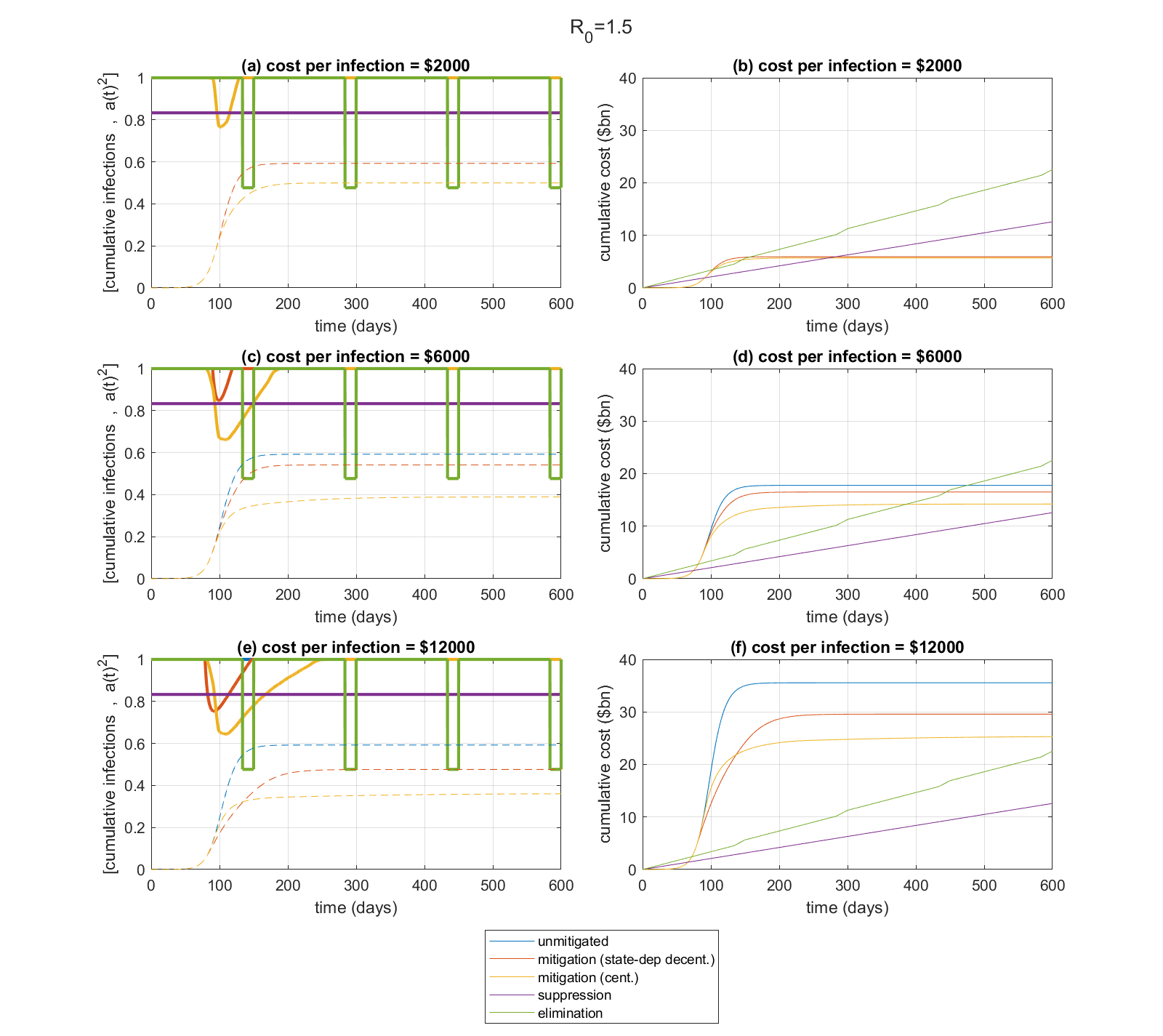}
    \caption{Comparison of the unmitigated (blue), decentralised mitigation with state-dependent behaviour (red), centralised mitigation (yellow), suppression (purple), and elimination (green) responses for $R_0=1.5$ and three values of the cost per infection $k$: (a,b) $k=$ \$2000; (c,d) $k=$ \$6000; (e,f) $k=$ \$12000. Left-hand plots (a,c,e) show the time-dependent transmission rate $a(t)^2$ relative to the unmitigated case (solid curves) and the cumulative proportion of the population infected (dashed curves). Note that infections are assumed to be negligible for suppression and elimination. Right-hand plots (b,d,f) show the cumulative cost. Parameter values as in Table 1. }
    \label{fig:state_dep1}
\end{figure}

\begin{figure}
    \centering
    \includegraphics[trim={2cm 0 2cm 0},clip,width=\linewidth]{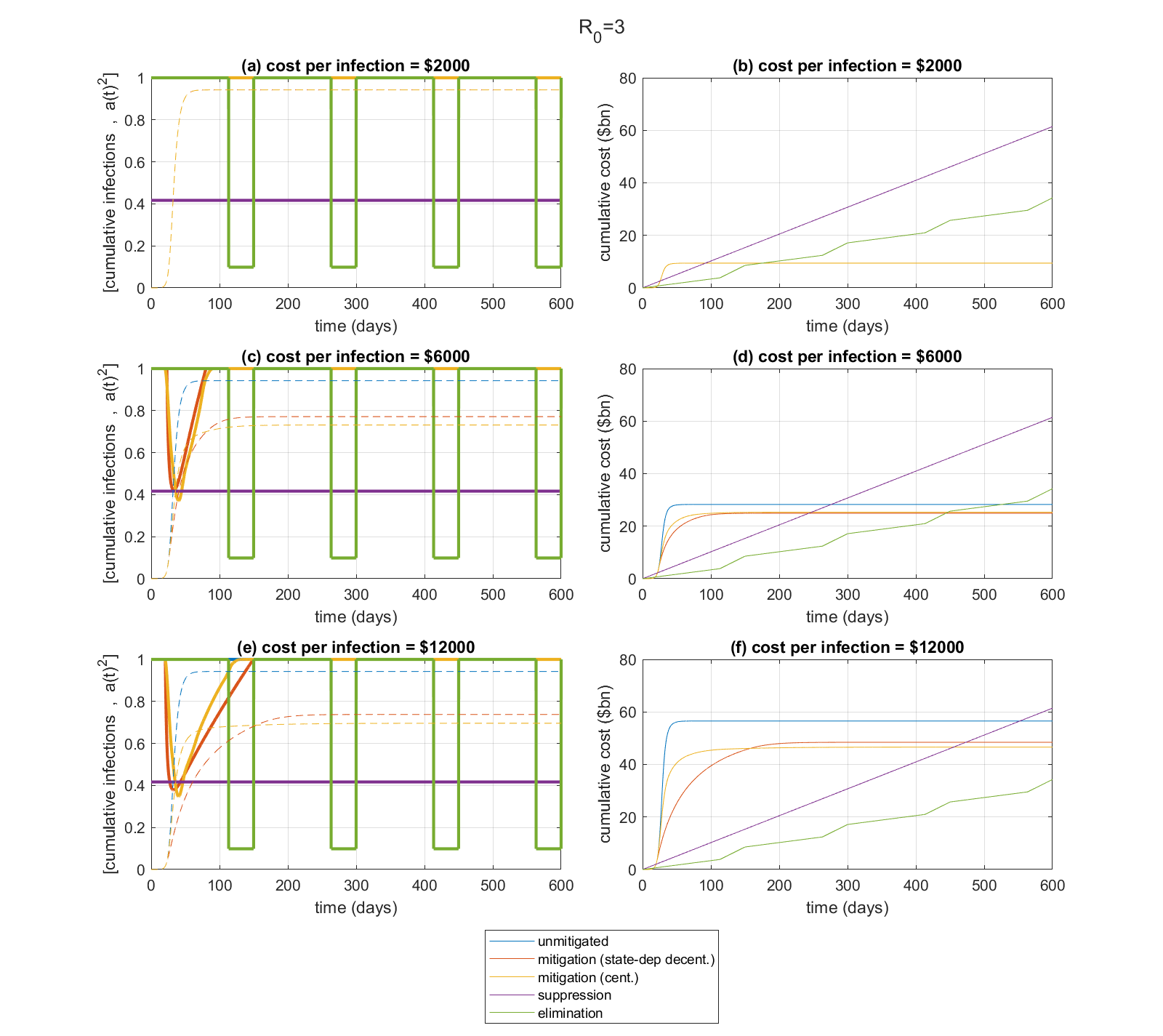}
    \caption{Comparison of the unmitigated (blue), decentralised mitigation with state-dependent behaviour (red), centralised mitigation (yellow), suppression (purple), and elimination (green) responses for $R_0=1.5$ and three values of the cost per infection $k$: (a,b) $k=$ \$2000; (c,d) $k=$ \$6000; (e,f) $k=$ \$12000. Left-hand plots (a,c,e) show the time-dependent transmission rate $a(t)^2$ relative to the unmitigated case (solid curves) and the cumulative proportion of the population infected (dashed curves). Note that infections are assumed to be negligible for suppression and elimination. Right-hand plots (b,d,f) show the cumulative cost. Parameter values as in Table 1. }
    \label{fig:state_dep2}
\end{figure}

\begin{figure}
    \centering
    \includegraphics[width=\linewidth]{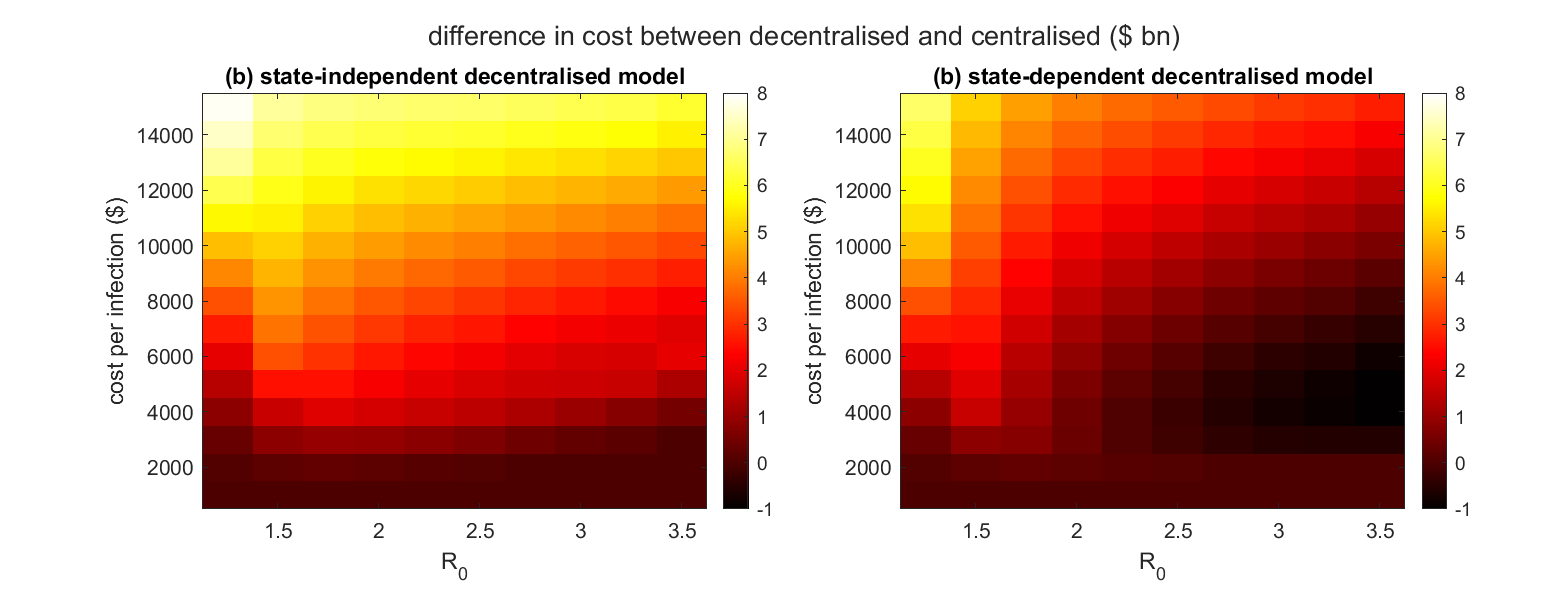}
    \caption{Difference in cost (\$ bn) between the decentralised and centralised mitigation responses for a range of values of $R_0$ and cost per infection $k$: (a) with the state-independent decentralised model; (b) with the state-dependent decentralised model. Other parameters as shown in Table 1.  }
    \label{fig:cost_diff}
\end{figure}

Results for the state-dependent decentralised model are shown in Figures \ref{fig:state_dep1} and \ref{fig:state_dep2} for $R_0=1.5$ and $R_0=3$ respectively. The state-dependent model produces a larger activity-level reduction than the equivalent state-independent model (Figures 1--2 of the main article), with an aggregate cost that is closer to (and in some cases slightly below) that of the centralised mitigation model (see Figure \ref{fig:cost_diff}).

\section{Infection costs under the elimination strategy}

The elimination strategy model assumes that border-related outbreaks occur with some frequency $r$ and grow undetected for time $t_\mathrm{det}$ before being controlled and eliminated by PHSMs. For simplicity, we ignore the costs associated with infections that occur in these border-related outbreaks. 
An approximate expression for the number of infections $n_\mathrm{outbreak}$ that are expected to occur in such an outbreak can be obtained by integrating the expected model incidence of infections per unit time $\gamma I(t)$ in the growth and control phases of the outbreak
\begin{equation}
    n_\mathrm{outbreak}= \gamma \left[ \int_0^{t_\mathrm{det}} e^{(\beta-\gamma)t} dt + \int_0^{t_c} e^{(\beta-\gamma)t_\mathrm{det}} e^{(\beta \alpha a_\mathrm{elim}^2-\gamma)t}dt \right]
\end{equation}
Evaluating the integrals and using Eq. (11) for the duration of control $t_c$ gives
\begin{equation}
    n_\mathrm{outbreak} = \frac{\gamma\left( e^{(\beta-\gamma)t_\mathrm{det}}-1\right)}{\gamma -\beta \alpha a_\mathrm{elim}^2}
\end{equation}
Finally, multiplying by the outbreak frequency $r$, the time horizon $T$ and the cost per infection $k$ gives the expected cumulative cost of infections $C_\mathrm{inf}$ occurring in border-related outbreaks up to time $T$:
\begin{equation}
    C_\mathrm{inf} = \frac{rkT \gamma\left( e^{(\beta-\gamma)t_\mathrm{det}}-1\right)}{\gamma -\beta \alpha a_\mathrm{elim}^2}
\end{equation}
We calculated $C_\mathrm{inf}$ for the range of values of $R_0$ and $k$ shown in Figure 3 of the main text. In all cases $C_\mathrm{inf}$ was less than 0.2\% of the total cost of the elimination response.

\section{Parameter values for Covid-19 in New Zealand}

New Zealand adopted an elimination strategy in response to the Covid-19 pandemic, using strong border controls and relying on reactive public health and social measures (PHSMs) to eliminate the virus when border incursions occurred \cite{baker2023covid}. In late 2020 the New Zealand Treasury estimated the daily economic cost of the controls used that year at the national and regional level \cite{TreasuryEconomicUpdateSept2020}. These controls were communicated to the public as a series of 'alert levels', ranging from alert level 1, which involved border controls but no domestic restrictions, through to alert level 4, which involved border controls and stringent school and business closures and stay-at-home orders \cite{baker2020successful}. The cost of alert level 1 represents the direct reduction in economic activity from border controls, while the costs of alert level 2 through 4 contain the additional losses of economic activity from domestic restrictions \cite{CBEHendy2025}. The estimates of $R_\mathrm{eff}$ and the corresponding economic costs at each alert level are shown in Table~\ref{tab:costs}.

\begin{table}[]
    \centering
    \begin{tabular}{cccc}
    \hline
       Control  & $R_\mathrm{eff}$ & $a_\mathrm{eff}$ & Cost \\
         &  & & (million \$/day) \\
       \hline
       Alert level 1 & 2.5 & 1.00 & 33.8\\
       Alert level 2 & 1.8 & 0.95 & 59.1 \\
       Alert level 3 & 0.75 & 0.61 & 143 \\
       Alert level 4 & 0.35 & 0.42 & 227\\
         \hline
    \end{tabular}
    \caption{Estimated effectiveness \cite{hendy2021mathematical} and daily economy-wide costs \cite{CBEHendy2025} of New Zealand's centralised Covid-19 control framework in 2020. The activity level $a_\mathrm{eff}$ for Alert Levels 2--4 was calculated via $R_\mathrm{eff}/R_0 = \alpha a_\mathrm{eff}^2$, assuming $R_0=2.5$ and TTI effectiveness of 20\% ($\alpha=0.8$). All costs are in New Zealand dollars (NZD).}
    \label{tab:costs}
\end{table}

We compute the corresponding activity, $a_\mathrm{eff}$ in Table~\ref{tab:costs}, by assuming that some of the reduction in $R_\mathrm{eff}$ comes from TTI measures and the remainder from population-wide reductions in activity. Here we assume TTI measures reduce $R_\mathrm{eff}$ by 20\%. At times New Zealand's trace, test, and isolate system probably performed more effectively than this but there were also periods where it was overwhelmed \cite{james2021successful}. With this assumption, we fit the cost function $c_1 \left(1-a\right) + c_2 \left(1-a\right)^2$ to the national costs of domestic restrictions (i.e. subtracting border control costs from the costs in Table~\ref{tab:costs}) at each activity level, we find that $c_1 =$ \$226 million per day and $c_2 =$ \$179 million per day (equivalent to \$45 per person per day and \$36 per person per day). Similarly, we estimate that $b$, the economic cost of border controls, is that of alert level 1 or \$33.8 million per day. All costs are in New Zealand dollars (NZD).

It is harder to directly estimate the cost per infection for an individual in New Zealand as very few people were exposed until 2022 \cite{vattiato2022modelling}. Instead we use data from South Korea, which relied principally on testing and contact tracing to control its first wave of COVID-19 infections and only one region, Daegu-Gyeongbuk, experienced a significant number of infections in the Spring of 2020. This allowed Aum et al \cite{AUM2021101993} to use a difference-in-differences approach to estimate that an increase of 1 infection per 1000 people in Daegu-Gyeongbuk caused employment in the region to fall by 2 to 3\%. If we take this as an estimate of the reduction in activity of individuals in response to an increase in infections, we find that $\Delta a/a$ is between $-20 \Delta i$ and $-30\Delta i$, where $i=I/N$ is the per capita infection prevalence. Differentiating Eq. (9) of the main text, which gives the activity level $a^*$ at the Nash equilibrium for decentralized control, with respect to $i^*$ we obtain the following:
\begin{equation}
\left(\frac{1}{a^*}\right)^2 \frac{da^*}{di^*}= -\frac{k \beta S_{i,\infty}^*}{c_1+2 c_2} 
\end{equation}
For $a^* \sim 1$, this suggests that $k \beta S_{i,\infty}^*$ is between $20\left(c_1+2c_2\right)$ and $30 \left(c_1+2 c_2\right)$ for the Daegu-Gyeongbuk region.

If we assume that, in the absence of strong central controls, New Zealanders would have responded to a wave of infections similarly to their South Korean counterparts, then we can use this relation to estimate a revealed cost per infection: $k \sim$ \$7000 to \$11000 using $c_1$ and $c_2$ from above. In other words, we estimate that in a counterfactual outbreak where no population wide controls had been imposed centrally, New Zealanders would have reduced their economic activity in a trade-off with a cost of infection of the order of between \$7000 and \$11000. 

An alternative approach to estimating the average cost per infection is to directly estimate the economic cost of Covid-19 fatalities, hospitalisations, and lost productivity due to people taking time off work while sick. Table \ref{tab:infection_cost} shows an example of such a calculation, based on low and high cost estimates. The cost of a Covid-19 fatality is the mean number of years of life lost (YLL) per Covid-19 death, estimated to be between 8 and 10 \cite{williams2022years,milkovska2025quantifying}, multiplied by an estimate of the cost of a disability-adjusted life year (DALY) in New Zealand of around \$94,000 \cite{daroudi2021cost,mbie_gdp}. This ignores the non-fatal health burden of Covid-19, although this is reasonable as it has been estimated that over 99\% of DALYs lost due to Covid-19 were due to fatalities \cite{mcdonald2022estimated}. 

The cost of a Covid-19 hospitalisation is the average cost of a hospital bed per night in New Zealand of \$1200 \cite{moh2024budget} multiplied by the mean length of stay for Covid-19 hospital patients of 8-9 days \cite{vekaria2021hospital}. Estimates for the infection-fatality ratio and infection-hospitalisation ratio in the 2020 pre-vaccine era were taken from \cite{ward2024real}. 

The productivity cost per infection was based on the Statistics NZ estimate for the mean annual income per person in New Zealand in 2020 (\$105,701) and an estimate for the proportion of SARS-CoV-2 infections resulting in clinical symptoms of 60--70\% \cite{sah2021asymptomatic}. We assumed that the mean time an individual was unable to work per symptomatic infection was 10 days. Overall, this leads to an estimated average cost per infection between around \$6500 and \$12000 (see Table \ref{tab:infection_cost}), which is a very similar range to the estimate based on revealed costs. This calculation ignores other costs such as the cost of primary (non-hospital) healthcare and the post-acute health burden.

\begin{table}
  \centering
    \begin{tabular}{p{7cm}rrp{2cm}}
    \hline   
    \textbf{Quantity} & \multicolumn{1}{l}{\textbf{Low estimate}} & \multicolumn{1}{l}{\textbf{High estimate}} & \multicolumn{1}{l}{\textbf{Source}} \\
    \hline    
    Infection-fatality ratio & 0.7\% & 1.2\% & \cite{ward2024real}  \\
    Years of life lost per Covid death & 12    & 14    & \cite{williams2022years} \\
    RLE reduction factor for Covid deaths & 0.72  & 0.72  & \cite{milkovska2025quantifying} \\
    GDP per capita & \$64,134 & \$64,134 & \cite{mbie_gdp} \\
    (DALY cost) / (GDP per capita) & 1.46  & 1.46  & \cite{daroudi2021cost} \\
    \textbf{Fatality cost per infection} & \textbf{\$5,663} & \textbf{\$11,326} &  \\
    \hline    
    Infection-hospitalisation ratio & 2.0\% & 3.0\% & \cite{ward2024real} \\
    Cost per bed night & \$1,200 & \$1,200 & \cite{moh2024budget} \\
    Mean length of hospital stay (days) & 8     & 9     & \cite{vekaria2021hospital} \\
    \textbf{Hospitalisation cost per infection} & \textbf{\$192} & \textbf{\$324} &  \\
    \hline   
    Mean annual household income & \$105,701 & \$105,701 & \cite{statsnz_household_income} \\
    People per household & 2.7   & 2.7   & \cite{statsnz_household} \\
    Clinical fraction & 60\%  & 70\%  & \cite{sah2021asymptomatic} \\
    Mean time unable to work per clinical infection (days) & 10     & 10     & Assumed \\
    \textbf{Productivity cost per infection} & \textbf{\$644} & \textbf{\$751} &  \\
    \hline
    \textbf{Total cost per infection} & \textbf{\$6,499} & \textbf{\$12,401} &  \\
    \hline
    \end{tabular}%

      \caption{Estimation of the average cost per SARS-CoV-2 infection in New Zealand in 2020 based on health and economic data. All costs are in New Zealand dollars. Notes: `RLE reduction factor for Covid deaths' is a multiplicative factor that was applied to account for the fact that the remaining life expectancy (RLE) for individuals who died from Covid-19 was estimated to be 28\% lower than standard life tables would indicate due to the contribution of comorbidities to the Covid-19 fatality risk \cite{milkovska2025quantifying}; where the source provides time-varying estimates, values from 2020 were used.  }
  \label{tab:infection_cost}%
\end{table}%

We estimate the frequency of border-related Covid-19 outbreaks in New Zealand during the elimination phase. Grout et al. \cite{grout2021failures} documented ten border failures in just over a year of operation of the managed isolation and quarantine system for international arrivals in 2020 to 2021. However, the majority of these were small clusters that were contained with TTI measures. Between April 2020, when the MIQ system was introduced, and October 2021 when the elimination strategy was ended, there were three outbreaks requiring the introduction of stringent PHSMs (alert level 3 or 4) in August 2020, February 2021, and August 2021. This translates to a frequency $r$ of approximately 1 per 150 days. The outbreaks were controlled largely by PHSMs in the Auckland region, which accounts for approximately 40\% of New Zealand's economy.  We therefore set $p_\mathrm{cont}=0.4$. We assume that border-related outbreaks are detected on average after $t_\mathrm{det}= 14$ days of uncontrolled community transmission. These are indicative values and we investigate the effect of varying some of these parameters in Results.

\bibliography{references}